\documentclass[aps,prb,amsmath,twocolumn]{revtex4}
\usepackage{bm}
\usepackage{graphicx}

\DeclareMathOperator{\sgn}{sgn}

\begin{document}

\title{Fractional Shapiro steps without fractional Josephson effect}
\author{Artem V. Galaktionov}
\affiliation{I.E. Tamm Department of Theoretical Physics, P.N. Lebedev Physical Institute, 119991 Moscow, Russia}
\author{Andrei D. Zaikin}
\affiliation{Institute for Quantum Materials and Technologies, Karlsruhe Institute of Nanotechnology (KIT), 76021, Karlsruhe, Germany}
\affiliation{I.E. Tamm Department of Theoretical Physics, P.N. Lebedev Physical Institute, 119991 Moscow, Russia}

\date{\today}
\begin{abstract}
It is widely believed that superconducting junctions involving topological insulators and hosting Majorana-like bound states may exhibit
unusual "fractional" ($4\pi$-periodic) ac Josephson effect.  Accordingly, "fractional" Shapiro steps on the current-voltage characteristics of such junctions are expected to occur under external microwave radiation. Here, we microscopically evaluate Shapiro steps in topologically trivial highly transparent superconducting weak links.  The key features recovered within our analysis -- including, e.g., the so-called "missing" Shapiro steps -- turn out to be similar to those observed in topological Josephson junctions. Our results demonstrate that caution is needed while interpreting experimental results for superconducting weak links in terms of Majorana physics.
\end{abstract}
\pacs{}
\maketitle
\section{Introduction and motivation}

Over recent years a great deal of attention was paid to experimental and theoretical investigations of superconducting junctions involving topological insulators which can host non-trivial gapless bound states. Such Majorana-like states are believed to exhibit rather exotic physical properties and are also regarded as promising in the context of topologically protected quantum computation \cite{AK}.

Theoretically a doublet of Majorana-like bound states was identified in point contacts formed by $p$-wave superconductors \cite{Yakovenko} as well as in junctions composed of two conventional ($s$-wave) superconductors connected to each other via two- and three-dimensional topological insulators \cite{FK,3D}. In all these cases the energies of these Majorana-like modes $E^M_{\pm}$ depend on the superconducting phase difference $\varphi$ across the junction as \cite{Yakovenko,FK,3D}
\begin{equation}
E^M_{\pm} (\varphi) = \pm\delta \cos \frac{\varphi}{2}.
\label{EM}
\end{equation}
For reasons and under conditions to be specified below it is widely believed that junctions hosting such Majorana-like modes exhibit an unusual $4\pi$-periodic current-phase relation (CPR) \cite{Yakovenko,FK} $I_M \propto \sin \frac{\varphi}{2}$ which is fundamentally different from the standard $2\pi$-periodic Josephson CPR $I_J=I_c\sin \varphi$. If so, one is bound to conclude that -- having in mind the Josephson relation $\dot\varphi =2eV$ (where $e$ defines electron charge and $V$ stands for bias voltage) -- the supercurrent oscillation frequency in such junctions equals to $eV$, i.e to a half of the standard Josephson frequency $\omega_J=2eV$.

 The presence of Majorana-like bound states \eqref{EM} could then be detected experimentally, e.g., by observing resonances in the form of current jumps (the so-called Shapiro steps \cite{Tinkh,BP,Likh}) on the the junction $I-V$ curve under the influence of external microwave radiation. Pronounced Shapiro steps at frequencies $\omega =\omega_J/2$ were indeed detected in a number of microwave experiments performed with HgTe- and BiSb-based superconducting junctions  \cite{M16,topmat,Gre}.

Very generally, internal properties of any weak link between two conventional superconductors -- no matter how complicated these properties are -- may influence its critical current and the form of CPR but {\it not} its periodicity in $\varphi$. The latter, in turn, is fundamentally determined by the periodicity of the charge space \cite{SZ90} reciprocal to the phase one. At subgap energies and temperatures the charge space of our problem remains effectively $2e$-periodic which immediately enforces $2\pi$-periodic in $\varphi$ CPR.

In the light of this, $4\pi$-periodic CPR could be regarded as quite unusual since in that case the charge transfer between superconducting condensates on both sides of the junction should be provided by quasiparticles with charge $e$ rather than by Cooper pairs with charge $2e$.
It remains unclear to us which physical mechanism could be responsible for such kind of charge transfer between two conventional superconductors.

For completeness, we also remark that general effective actions \cite{SZ90,Z94,SN,book} describing superconducting junctions with arbitrary
transmissions do, of course, include $4\pi$-periodic terms in the Josephson phase $\varphi$. However, such (non-local in time) terms account for {\it dissipative} currents and have nothing to do with the supercurrent flowing across the weak link.

Below we will consider a purely ballistic superconducting weak link (or an SNS junction) with ${\mathcal N}$ fully transparent conducting channels and normal state conductance $1/R_N={\mathcal N}e^2/\pi$. Provided the thickness of a normal (N) layer $d$ connecting two superconducting (S) electrodes  is much smaller than the coherence length $d \ll \xi_0 \sim v_F/\Delta$ the junction hosts a pair of subgap Andreev bound states
\begin{equation}
E^A_{\pm} (\varphi) = \pm\Delta \cos \frac{\varphi}{2}
\label{EA}
\end{equation}
in each of ${\mathcal N}$ conducting modes, see also Fig. 1. Here and below $\Delta$ stands for the superconducting order parameter in the electrodes and $v_F$ is the Fermi velocity. Since the phase dependence of the bound states energy \eqref{EA} is identical to that for Majorana-like states \eqref{EM}, ballistic SNS junctions can serve as a convenient playground to test some ideas and approaches to superconducting junctions based on topological insulators.
\begin{figure}
\includegraphics[width=7cm]{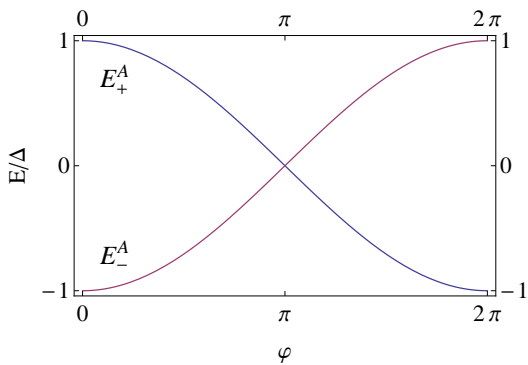}
\caption{A pair of Andreev bound states $E^A_{\pm} (\varphi)$ in short ballistic SNS junctions.}
\label{Fig1}
\end{figure}

In equilibrium CPR of short ballistic SNS junctions takes the well-known $2\pi$-periodic in $\varphi$ form \cite{KO}
\begin{equation}
I=I_S(\varphi) =\frac{\pi\Delta}{eR_N}\sin\frac{\varphi}{2}\tanh\left(\frac{\Delta}{2T}\cos\frac{\varphi}{2}\right).
\label{sjc}
\end{equation}
Exactly the same relation also holds for short symmetric SINIS junctions at resonance for arbitrary (including very low)
transmissions of SN interfaces \cite{GZ02}. The relation \eqref{sjc} can also be recovered from a simple formula \cite{FT,BH}
\begin{equation}
I=2e{\mathcal N}\sum_{\pm}\frac{\partial E^A_\pm}{\partial \varphi}f_\pm,
\label{dEdf}
\end{equation}
where $f_\pm\equiv f_F(E^A_\pm)=1/[1+\exp (E^A_\pm/T)]$ are the Fermi filling factors \cite{FN} for both Andreev bound states. This formula indicates that equilibrium CPR in weak links with $d \ll \xi_0$ can be associated only with discrete Andreev levels \eqref{EA}, whereas continuous electron spectrum does not play any role in this limit.

Making use of this observation, the same formula with $E^A_\pm\to E^M_\pm$ was employed \cite{Yakovenko,FK,3D} in order to identify the contribution from Majorana-like states \eqref{EM} to the supercurrent flowing across the junctions involving topological insulators. In equilibrium this contribution is again $2\pi$-periodic in $\varphi$, just like CPR in Eq. \eqref{sjc}.

One may be tempted to extend the formula \eqref{dEdf} in order to describe ac Josephson effect. Following \cite{Yakovenko}, by sweeping the phase $\varphi$ sufficiently slowly, but not too slowly, one would be able to drive the system in such a way that it has no time to relax to its ground state, thereby always occupying one of the states \eqref{EM} and leaving another one empty. Assuming  $\delta < \Delta$ (which typically requires violation of time-reversal symmetry) and $eV < \Delta -\delta$ in order to separate Majorana-like levels from the continuum of states \cite{FK} and substituting the filling factors $f_-=1$ and $f_+=0$ into Eq.  \eqref{dEdf} with $E^A_\pm\to E^M_\pm$ one immediately arrives at the contribution to the supercurrent in the desired form $I_M \propto \sin (\varphi (t)/2)$.

In our view, this procedure cannot yet be regarded as a rigorous derivation and in any case it should be verified by means of an elaborate quantum kinetic analysis. This analysis should fully account, e.g., for the effect of multiple Andreev reflection (MAR) \cite{MAR} which is known to play a crucial role in junctions with not very low transmissions. Fortunately in the case of ballistic SNS junctions considered here such detailed microscopic theory is already well established \cite{Zaitsev80,Zaikin83,Uwe,AB1,AB2,Uwe2}.  This theory demonstrates that MAR serves as a key charge transfer mechanism across such junctions setting in already at arbitrarily small bias voltages $V=\dot\varphi/2e$.

\begin{figure}
\includegraphics[width=7cm]{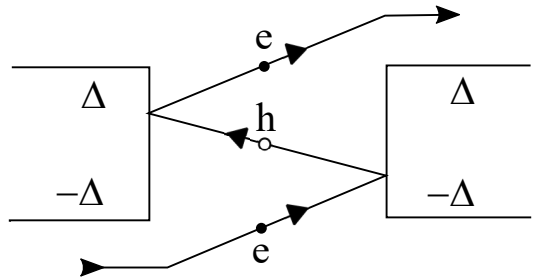}
\caption{Schematics of multiple Andreev reflection process in ballistic SNS junctions.}
\label{Fig2}
\end{figure}

The process of multiple Andreev reflection is schematically illustrated in Fig. 2. After each traverse across the junction a quasiparticle (hole) gets accelerated by $eV$, suffers Andreev reflection at one of the two NS interfaces and eventually leaves the junction after $m \approx 2\Delta /eV$ such traverses. As a result, the distribution functions $f^+(E)$ and $f^-(E)$ for quasiparticles moving respectively in and opposite to the current direction deviate strongly from the equilibrium distribution function $f_F(E)$. They read (see, e.g., Ref. \onlinecite{Uwe2})
\begin{equation}
f^{\pm}(E)=\sum_{m=0}^{\infty}f_F(E\mp meV)[1-{\mathcal A}(E\mp meV)]\prod_{l=0}^{m-1} {\mathcal A}(E\mp leV),
\nonumber
\end{equation}
where ${\mathcal A}(E)$ is the Andreev reflection probability. At $T, eV \ll \Delta$  one arrives at the $I-V$ curve in the form  \cite{Uwe}
\begin{equation}
\bar{I} = \frac{V}{R_N}+ \frac{2\Delta }{eR_N}\sgn V,
\label{IV1}
\end{equation}
where $\sgn x$ equals to 1, 0 and -1 respectively for $x>0$, $x=0$ and $x<0$. The last term in Eq. \eqref{IV1} represents an excess current due to MAR and holds at arbitrary values of $d$. In the absence of inelastic relaxation this fairly large current sets in at any non-zero applied voltage $V$.

In the limit of short SNS junctions one also finds \cite{AB1}
\begin{equation}
I(t)=\frac{V}{R_N}+I_c|\sin (\varphi /2)|\sgn V, \quad \varphi =2eVt,
\label{smallVIphi}
\end{equation}
where $I_c=\pi\Delta/(eR_N)$ is the junction critical current at $T \to 0$. We observe that CPR \eqref{smallVIphi} remains strictly $2\pi$-periodic and no $4\pi$-periodic supercurrent component occurs at non-zero voltages. Averaging Eq. \eqref{smallVIphi} over time, we again recover Eq. \eqref{IV1}.

As we already discussed, in the presence of external microwave radiation with frequency $\omega$  the junction $I-V$ curve exhibits Shapiro steps which occur under the condition \cite{AB2,Uwe2,Cuevas} $2keV=n\omega$ or, equivalently, at
\begin{equation}
\omega=\frac{k}{n}\omega_J,
\label{Shst}
\end{equation}
where $k$ and $n$ are positive integer numbers. For clarity, in what follows we will distinguish integer and fractional Shapiro steps corresponding to respectively integer and non-integer values of the ratio $k/n$ in Eq. \eqref{Shst}. The values $k\geq 2$ (irrelevant in the tunneling limit) reflect the presence of higher harmonics of the Josephson current due to non-sinusoidal CPR, whereas the numbers $n\geq 2$ correspond to multi-photon processes which gain importance at higher ac signals or smaller microwave frequencies.

Hence, fractional Shapiro steps are routinely expected even in topologically trivial Josephson junctions and the observation of such steps (e.g., at  $\omega =\omega_J/2$) alone is by no means sufficient to make any definite conclusion about  the presence of a $4\pi$-periodic in $\varphi$ contribution to CPR. In addition to fractional Shapiro steps, in experiments \cite{M16,topmat,Gre} missing integer Shapiro steps at  $\omega =\omega_J$ were reported at low enough frequencies whereas at higher $\omega$ these steps were clearly detected. With the aid of a phenomenological model it was argued \cite{4pisteps} that such observations actually support the $4\pi$-periodic Josephson effect scenario.
We will return back to this issue below in Section VIII.

The structure of the paper is as follows. In Section II we outline our general formalism which is then employed in Section III in order
to evaluate electric current across ballistic SNS junctions for a arbitrary time dynamics of the Josephson phase. In Section IV we reconstruct a complete description of ac Josephson effect in SNS junctions biased by a constant external voltage. The effect of an external ac signal on both CPR and the Josephson critical current is discussed in Section V. Sections VI and VII are devoted to the analysis of both integer and fractional Shapiro steps respectively in the voltage- and current-biased regimes. Discussion of our key observations is presented in Section VIII. Some technical details of our calculation are relegated to Appendices A, B, C and D.

\section{Basic formalism}
In what follows we will consider a short $SNS$ junction with $d \ll \xi_0$. This system can be conveniently described within the standard technique combining quasiclassical Eilenberger equations \cite{,Eil,ZRev} with Zaitsev boundary conditions  \cite{Zai} matching quasiclassical propagators on both sides of the junction. Making use of this approach it is straightforward to derive a general expression for the effective action \cite{Z94} that accounts for arbitrary configurations of the Josephson phase $\varphi (t)$ and holds for any junction transmission distribution.

As we are aiming at describing non-stationary and non-equilibrium  processes it is necessary to resort to the Keldysh technique and introduce
two phase variables $\varphi_{1}(t)$ and $\varphi_{2}(t)$ defined respectively on the forward and backward parts of Keldysh contour.
Then the Keldysh version of the effective action reads \cite{SN,book}
\begin{equation}
iS_t[\varphi]=\frac{1}{2}\sum_n {\rm Tr}\,\ln\left[ 1+\frac{{\mathcal T}_n}{4}\left(\left\{\check Q_L(\varphi),\check Q_R\right\}-2 \right)\right].\label{ea}
\end{equation}
Here the summation runs over the junction conducting channels with arbitrary transmission coefficients ${\mathcal T}_n$, $\check Q_{L,R}$ are $4\times4$ Green-Keldysh matrices for the left and right superconducting electrodes. The product of these matrices implies time convolution and curly brackets denote anticommutation. These matrices are expressed through quasiclassical propagators as
\begin{equation}
\check Q_R=\check g=\left( \begin{array}{cc} \hat g^R& \hat g^K\\0
& \hat g^A\end{array}\right)
\label{propagator}
\end{equation}
and
\begin{equation}
\check Q_L(\varphi)(t,t')=\check {\cal L} \check {\cal
M}_+(t)\check {\cal L} \check g(t,t') \check {\cal L} \check {\cal
M}_-(t')\check {\cal L}.
\end{equation}
The retarded, advanced and Keldysh components of the matrix \eqref{prop} are in turn $2\times2$ matrices in the Nambu space
\begin{equation}
\hat g^{R,A,K}(\epsilon)=\left(\begin{array}{cc} g^{R,A,K}(\epsilon) & f^{R,A,K}(\epsilon) \\ -f^{R,A,K}(\epsilon) & - g^{R,A,K} (\epsilon) \end{array}\right),
\end{equation}
and we also define the matrices
\begin{equation}
\check{\cal M}_{\pm}= \left(
\begin{array}{cc} \exp\left[\pm i\varphi_1(t)\hat\tau_3/2\right] & 0 \\  0 &
 \exp\left[\pm i\varphi_2(t)\hat\tau_3/2\right]\end{array}\right),
 \end{equation}
\begin{eqnarray}
\check {\cal L}=\frac{1}{\sqrt{2}}\left( \begin{array}{cc}\hat 1&
\hat 1\\ \hat 1&-\hat 1 \end{array}\right), \quad \hat\tau_3=\left(\begin{array}{cc} 1& 0\\ 0& -1 \end{array} \right).
\end{eqnarray}

The retarded and advanced propagators obey the following conditions:
\begin{eqnarray}
\label{prop}
\left(g^R(\epsilon)\right)^2-\left(f^R(\epsilon)\right)^2=1,\quad \lim_{\epsilon\rightarrow \pm \infty} g^R(\epsilon)=1,\\
g^R(-\epsilon)=(g^R(\epsilon))^*,\quad f^R(-\epsilon)=-(f^R(\epsilon))^*,
\label{prop2}\\
g^A(\epsilon)=-\left(g^R(\epsilon)\right)^*,\quad f^A(\epsilon)=-\left(f^R(\epsilon)\right)^*,\label{arr}
\end{eqnarray}
whereas the Keldysh components read
\begin{eqnarray}
g^K(\epsilon)= 2\, {\rm Re} \left[ g^R(\epsilon)\right] \tanh\frac{\epsilon}{2T}, \\
\quad f^K(\epsilon)= 2\, {\rm Re} \left[ f^R(\epsilon)\right] \tanh\frac{\epsilon}{2T}.\label{kc}
\end{eqnarray}
In the case of conventional superconducting electrodes we have
\begin{equation}
g^{R,A}(\epsilon)=\frac{\epsilon\pm i\gamma}{\xi^{R,A}(\epsilon)},\quad f^{R,A}(\epsilon)=\frac{\Delta}{\xi^{R,A}(\epsilon)},\label{arch}
\end{equation}
where $\xi^{R,A}(\epsilon)=\pm\sqrt{(\epsilon\pm i\gamma)^2-\Delta^2}$ and $\gamma$ controls the strength of inelastic relaxation. In the absence of inelastic relaxation it is necessary to keep $\gamma$ infinitesimally small.

In order to proceed let us introduce the "classical" and "quantum" phases in as standard manner as $\varphi_+=(\varphi_1+\varphi_2)/2$ and
$\varphi_-=\varphi_1-\varphi_2$, respectively. A general expression for the current operator expectation value can be defined in terms of the path integral
\begin{equation}
\langle \hat I(t)\rangle =2ie\int {\cal D} \varphi_{\pm}\frac{\delta}{\delta
  \varphi_-(t)}e^{iS_c[\varphi_\pm]+iS_t[\varphi_\pm]}.
 \label{curr}
\end{equation}
which should be evaluated under the conditions $\langle \dot\varphi_+(t)\rangle=2eV(t)$ and $\langle \dot\varphi_-(t)\rangle=0$, where $V(t)$ is the voltage across the junction. The expression \eqref{curr} can also be employed in the presence of electron-electron interactions in which case the charging term $S_c$ should be added to the effective action $S_t$ \eqref{ea}.

Perhaps we can add that the above effective action formalism is equally convenient for the analysis of higher cumulants of the current operator.
The current-current correlation function as well as higher current correlators are obtained by applying consecutive derivatives $\delta/\delta\varphi_-(t)$ under the path integral \eqref{curr}. In the particular case of relatively small phase fluctuations this procedure was implemented in Ref. \onlinecite{GZ}.

Here the "classical" phase variable $\varphi_+(t)=2 e\int^t dt'V(t')$ cannot anymore be regarded as small which makes the whole problem rather complicated to deal with. On the other hand, in the absence of electron-electron interactions the path integral in Eq. \eqref{curr} is evaluated trivially and we obtain
\begin{equation}
I(t)=\left. -2e\frac{\delta S_t[\varphi_\pm]}{\delta \varphi_-(t)}\right|_{\varphi_-=0,\; \varphi_+(t)\equiv \varphi (t)=2e\int^t dt'V(t') }.\label{cgf}
\end{equation}
From now on we will make no distinction between the phase variables $\varphi_+(t)$ and $\varphi (t)$.

\section{Electric current for ballistic junctions}

In the interesting for us limit ${\mathcal T}_n=1$ the action (\ref{ea}) becomes simpler. With the aid of the normalization condition $\check Q_L^2(t,t')= \check Q_R^2(t,t')=\delta(t-t')$, in this limit we have
\begin{eqnarray}
\nonumber
iS_t[\varphi]=\frac{1}{2}\sum_n {\rm Tr}\,\ln\left[ \frac{1}{4}\left( \check Q_L(\varphi)+\check Q_R\right)^2\right]\\={\mathcal N} {\rm Tr}\,\ln\left[ \frac{1}{2}\check I \left( \check Q_L(\varphi)+\check Q_R\right)\right], \label{eda}
\end{eqnarray}
where the matrix $\check I$ is defined as
\begin{equation}
\check I=\left(\begin{array}{cc} \hat\tau_3 & 0\\0 & -\hat\tau_3\end{array}\right).
\end{equation}
Expanding the argument of the logarithm in Eq. (\ref{eda}) up to the first order in $\varphi_-(t)$ and symmetrizing the resulting expressions with respect to the phase variables, we get
\begin{equation}
iS_t={\mathcal N}{\rm Tr}\,\ln\left[ \check Q_0+\check Q_1\right].
\label{eaexp}
\end{equation}
Here the matrices $\check Q_0$ and $\check Q_1$ read
\begin{eqnarray}
\label{Q0}
\check Q_0=\left( \begin{array}{cc} \hat a^R & \hat a^K \\ 0 & -\hat a^A \end{array}\right), \quad {\rm Tr}\,\ln\left[ \check Q_0\right]=0,\\
\nonumber
\check Q_1(t,t')=\frac{\varphi_-(t)}{8}\left(\begin{array}{cc} 0 & -\hat b^A(t,t')\\ \hat b^R(t,t') & \hat b^K(t,t') \end{array} \right)\\
+ \left(\begin{array}{cc} \hat \tau_3\hat b^K(t,t')\hat\tau_3 &  \hat \tau_3\hat b^R(t,t')\hat\tau_3 \\ -\hat \tau_3\hat b^A(t,t')\hat\tau_3 & 0\end{array} \right)\frac{\varphi_-(t')}{8},
\label{Q1}
\end{eqnarray}
where we define
\begin{widetext}
\begin{equation}
\hat a^{R,A,K}(t,t')=\left(\begin{array}{cc} g^{R,A,K}(t,t')\cos\left[\frac{\varphi (t)-\varphi (t')}{4} \right] & f^{R,A,K} (t,t')\cos\left[\frac{\varphi (t)+\varphi (t')}{4} \right] \\ f^{R,A,K}(t,t')\cos\left[\frac{\varphi (t)+\varphi (t')}{4} \right]& g^{R,A,K}(t,t')\cos\left[\frac{\varphi (t)-\varphi (t')}{4} \right]\end{array}\right),\label{adef}
\end{equation}
\begin{equation}
\hat b^{R,A,K}(t,t')=\left(\begin{array}{cc} g^{R,A,K}(t,t')\sin\left[\frac{\varphi (t)-\varphi (t')}{4} \right] & f^{R,A,K}(t,t')\sin\left[\frac{\varphi (t)+\varphi (t')}{4} \right] \\ f^{R,A,K}(t,t')\sin\left[\frac{\varphi (t)+\varphi (t')}{4} \right] & g^{R,A,K}(t,t')\sin\left[\frac{\varphi (t)-\varphi (t')}{4} \right]\end{array} \right).\label{bexp}
\end{equation}
\end{widetext}
Combining Eq. (\ref{cgf}) with Eqs. \eqref{eaexp}-\eqref{Q1}, we arrive at the general expression for the current
\begin{eqnarray}
\nonumber
I(t)=\frac{ie{\mathcal N}}{4}\int dt' {\rm Tr}\left[ \hat b^R(t,t')\hat X^K(t',t)+\hat b^K(t,t')\hat X^A(t',t)\right.\\
\left.-\hat X^K(t,t')\hat \tau_3 \hat b^A(t',t) \hat \tau_3
 +\hat \tau_3 \hat X^R(t,t') \hat \tau_3 \hat b^K(t',t)\right],\label{abstcu}
\end{eqnarray}
where
\begin{eqnarray}
\hat X^R=\left( \hat a^R\right)^{-1}, \quad \hat X^A=-\left( \hat a^A\right)^{-1}
\label{XRA}
\end{eqnarray}
and
\begin{eqnarray}
 \hat X^K=-\hat X^R\circ \hat a^K\circ \hat X^A
\label{XK}
\end{eqnarray}
are the matrix elements of the inverse matrix
\begin{eqnarray}
\check Q_0^{-1}\equiv \check X=\left(\begin{array}{cc}\hat X^R & \hat X^K \\ 0 & \hat X^A \end{array}\right).
\label{cX}
\end{eqnarray}
The expression \eqref{abstcu} has an explicit causal nature and remains valid for an arbitrary dependence of the applied voltage $V(t)$ on time.
The general result \eqref{abstcu} coincides with those derived previously \cite{Zaitsev80,Zaikin83} by directly solving the Eilenberger equations.

\section{Time-independent bias voltage}
Let us first consider the limiting case of a constant bias voltage $V$ applied directly to the junction. Obviously,
the Josephson phase then depends linearly on time, i.e. $\varphi (t)=2eVt$. As it was demonstrated in Ref. \onlinecite{Uwe}, in this particular case it is possible to explicitly invert the matrices in Eq. \eqref{XRA} and recover the exact expression for the current $I(t)$ across the junction \cite{Uwe,AB1}.

As compared to Ref. \onlinecite{Uwe}, here we pursue a different approach outlined in Appendix A. Introducing the notation
\begin{equation}
a^R(\epsilon)=\frac{f^R(\epsilon)}{1+ g^R(\epsilon) },
\label{are}
\end{equation}
which has to do with the so-called Riccati parametrization (see, e.g., Ref. \onlinecite{ZRev})
\begin{equation}
f^R=\frac{2 a^R}{1-(a^R)^2},\quad g^R=\frac{1+(a^R)^2}{1-(a^R)^2}, \label{rp}
\end{equation}
and employing the (corresponding to MAR) multiplicative structure of the resulting expressions (\ref{mel}), from Eq. (\ref{abstcu}) we  arrive at the final result
\begin{equation}
I(t)=\sum_{l=-\infty}^\infty I_l e^{-2ielVt},
\label{Fou}
\end{equation}
where
\begin{widetext}
\begin{eqnarray}
I_{l=0}\equiv \bar{I} =\frac{V}{R_N}-\frac{1}{eR_N}\sum_{n=1}^\infty\int\limits_{-\infty}^{\infty} d\epsilon \tanh\frac{\epsilon}{2T}\left(1- \left| a^R(\epsilon)\right|^2\right)\prod_{1\le m\le n}\left| a^R(\epsilon+meV)\right|^2\label{iv},\\
I_{l>0}= -\frac{1}{eR_N}\sum_{n=1}^\infty\int\limits_{-\infty}^{\infty} d\epsilon \tanh\frac{\epsilon}{2T}\left(1- \left| a^R(\epsilon)\right|^2\right)\prod_{1\le m\le n}\left| a^R(\epsilon+meV)\right|^2\prod_{n+1\le k\le n+2l} a^R(\epsilon+keV)\nonumber\\
-\frac{1}{eR_N}\int\limits_{-\infty}^{\infty} d\epsilon \tanh\frac{\epsilon}{2T}\left(1- \left| a^R(\epsilon)\right|^2\right)\prod_{1\le k\le 2l} a^R(\epsilon+keV),
\quad I_{-l}=I_l^*
\label{arharm}.
\end{eqnarray}
\end{widetext}
The results (\ref{Fou})-(\ref{arharm}) fully conform to those of Refs. \onlinecite {Uwe,AB1}. As expected, the current $I(t)$ \eqref{Fou} is strictly $2\pi$-periodic in $\varphi (t)=2eVt$ for any nonzero $V$ and no $4\pi$-periodic component occurs.

Equation (\ref{iv}) provides a general expression for the average current $\bar{I}$. As we already discussed, it is characterized by the current jump at $V=0$ caused by MAR. This excess current also persists at all larger voltages.  Provided inelastic relaxation effects remain weak and can be neglected, combining Eqs. \eqref{iv} and (\ref{arch}) we obtain \cite{Uwe}
\begin{equation}
\bar{I} = \frac{V}{R_N}+ \frac{2\Delta (T)}{eR_N}\tanh\frac{\Delta(T)}{2T}\sgn V, \quad eV \ll \Delta (T)
\label{IV10}
\end{equation}
and \cite{Zaitsev80,Zaikin83,MAR}
\begin{equation}
\bar{I} = \frac{V}{R_N}+ \frac{8\Delta (T)}{3eR_N}\tanh\frac{eV}{2T}, \quad eV \gg \Delta (T).
\label{excess}
\end{equation}
Temperature effects do not eliminate the current jump although diminish its amplitude. However, non-vanishing inelastic relaxation, if present, smears this jump making the linear junction conductance $\bar{I}/V$ finite in the limit $V\to 0$ \cite{Uwe}.

At small voltages and temperatures $eV, T \ll \Delta$ and for weak inelastic relaxation Eqs. (\ref{Fou})-(\ref{arharm}) can be combined reducing to a simple expression for the total current defined in Eq. \eqref{smallVIphi}. We will explicitly make use of this result further below. Finally, at high voltages $eV \gg \Delta$ and low $T$ one readily finds \cite{AB1}
\begin{equation}
I(t) \simeq \bar{I}-\frac{\pi \Delta^2 \ln 2}{2e^2VR_N}\cos (2eVt).
\label{largeVIphi}
\end{equation}

\section{Supercurrent under ac signal}
As a next step we consider a somewhat different physical situation assuming now that our SNS junction is exposed to external microwave radiation which generates an ac voltage signal
\begin{equation}
V(t)=V_{\rm ac}\cos (\omega t)
\label{vac}
\end{equation}
across the junction. Here $V_{\rm ac}$ scales with the intensity of external radiation and $\omega$ is the radiation frequency. The Josephson phase then oscillates in time and takes the form
\begin{equation}
\varphi (t) =\bar \varphi +2\alpha \sin (\omega t), \quad \alpha=eV_{\rm ac}/\omega .
\label{phase}
\end{equation}

In this state with zero average voltage $\bar V$ the junction can still carry a non-vanishing dc supercurrent which is now affected by external radiation.  In order to evaluate this supercurrent one can employ the general formula (\ref{abstcu}) combined with Eq. \eqref{phase}. However, in contrast to the case of a constant in time bias voltage considered above, here it is not possible to exactly invert the corresponding matrices in Eq. (\ref{abstcu}) and one should resort to certain approximations. Physically the main complication is that in the presence of an external ac field the quasiparticle distribution function in the contact area is driven out of equilibrium an in a general case can only be evaluated numerically by resolving a self-consistent quantum kinetic equation which simultaneously accounts for Andreev reflection as well as photon absorption and emission processes. The corresponding analysis was carried out in Ref. \onlinecite{Uwe2}. The influence of external radiation on both CPR and the critical Josephson current in superconducting point contacts was previously studied in Ref. \onlinecite{Bergeret}.

A substantial simplification of the problem can be achieved if the supercurrent is evaluated within the adiabatic approximation which we will employ here. Assuming that the frequency $\omega$ is small enough to obey the condition
\begin{equation}
\omega \ll 2|E^A_{\pm} (\bar \varphi)|,
\label{adco}
\end{equation}
and neglecting (typically rather small) geometric capacitance of our junction one finds the time-dependent current $I(t)$ which reads \cite{GZ}
\begin{equation}
I(t)\simeq I_S(\varphi (t))+C^*(\bar\varphi )\ddot\varphi (t),
\label{acI}
\end{equation}
where $I_S(\varphi )$ is defined in Eq. \eqref{sjc} and $C^*(\bar\varphi )$ represents the renormalized junction capacitance. In the low temperature limit it takes the form \cite{GZ}
\begin{equation}
C^*=\frac{\pi}{16\Delta R_N\cos^4 (\bar\varphi /4)}
\label{rc}
\end{equation}
outside an immediate vicinity of the point $\bar\varphi = \pi$ and formally diverges at  $\bar\varphi \to \pi$, thus signaling the failure of the
adiabatic approximation for $E^A_{\pm} (\pi) =0$.

\begin{figure}
\includegraphics[width=8cm]{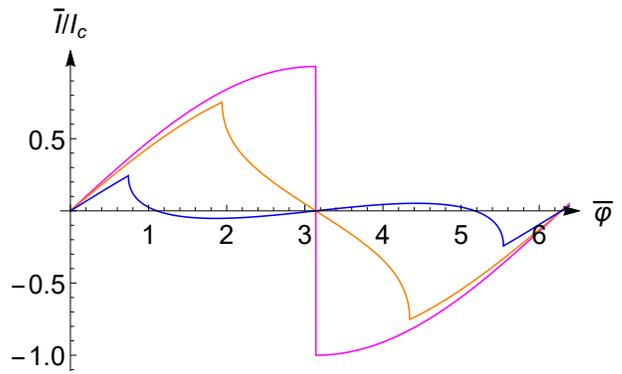}
\caption{Zero temperature current-phase relation $\bar I (\bar \varphi )$ (normalized by $I_c\equiv I_c(0)$) for $\alpha =0$ (magenta), 0.6 (orange) and 1.2 (blue).}
\label{Fig5}
\end{figure}

\begin{figure}
\includegraphics[width=8cm]{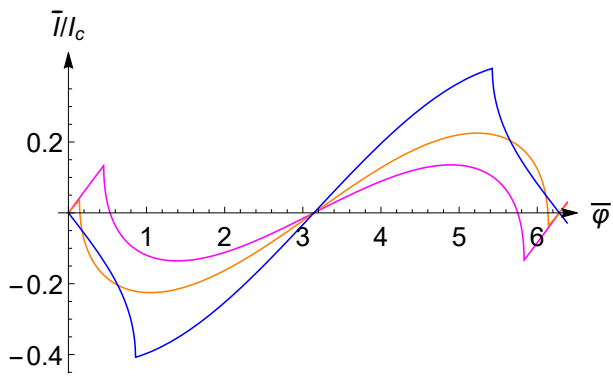}
\caption{The same as in Fig. \ref{Fig5}  for $\alpha =1.345$ (magenta), 1.5 (orange) and 2.0 (blue)(blue).}
\label{Fig6}
\end{figure}

\begin{figure}
\includegraphics[width=8cm]{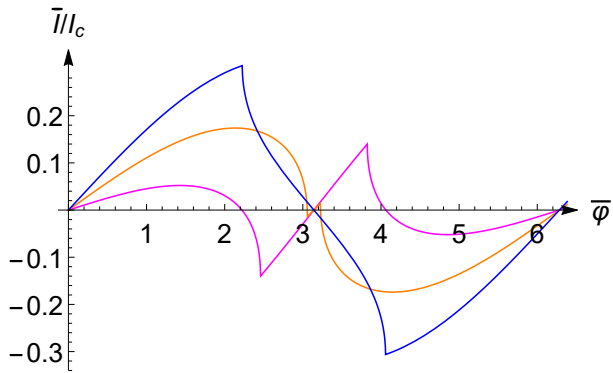}
\caption{The same as in Fig. \ref{Fig5}  for $\alpha =2.8$ (magenta), 3.1 (orange) and 3.6 (blue).}
\label{Fig7}
\end{figure}

\begin{figure}
\includegraphics[width=8cm]{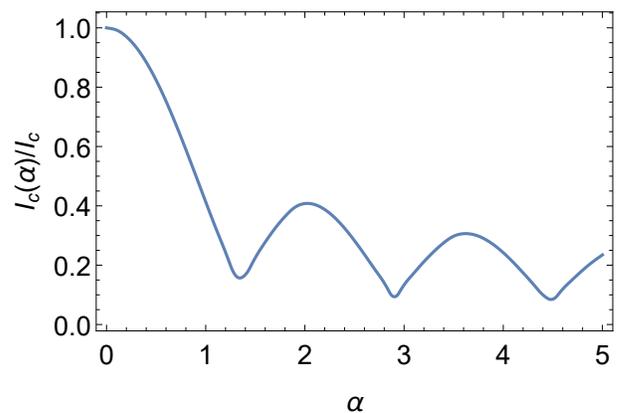}
\caption{Zero temperature critical current $I_c(\alpha )$ normalized by $I_c\equiv I_c(0)$. The current  $I_c(\alpha )$ remains non-zero at any finite value of $\alpha$. Local minima of this dependence correspond to concequtive transitions between 0- and $\pi$-junction states.}
\label{Fig8}
\end{figure}

Substituting the phase $\varphi (t)$ in the form  \eqref{phase} into Eq. (\ref{acI}), making use of Eq. \eqref{sjc} and averaging the resulting
expression over time, in the limit  $T \to 0$ we obtain an obvious relation (cf. also \cite{Bergeret})
\begin{equation}
\bar I(\bar\varphi ) = \frac{8\Delta}{eR_N}\sum_{m=1}^{\infty}\frac{m(-1)^{m+1}}{4m^2-1}J_0(2m\alpha )\sin (m\bar\varphi ),
\label{CPRac}
\end{equation}
where $J_0(x)$ is the zero-order Bessel function. For $\alpha < \pi/2$ there exists the phase interval $2\alpha -\pi \leq  \bar\varphi \leq  \pi - 2\alpha$ where the series \eqref{CPRac} can be summed up exactly with the result
\begin{equation}
\bar I(\bar\varphi ) = I_c J_0(\alpha ) \sin (\bar \varphi /2).
\label{CPRa}
\end{equation}
Outside this interval CPR $\bar I(\bar\varphi )$ \eqref{CPRac} can be evaluated numerically. The corresponding dependencies are illustrated in Figs. \ref{Fig5}-\ref{Fig7} for different values of $\alpha$.
We observe that Eq. \eqref{CPRa} is fully confirmed under its validity conditions, whereas beyond them CPR develops qualitatively new features including, e.g., negative current states within the interval $0< \bar\varphi < \pi$. For $\alpha \lesssim 1.345$ CPR reaches its absolute maximum at $ \bar\varphi = \pi -2\alpha$ in which case the junction critical current $I_c(\alpha )\equiv {\rm max} |\bar I(\bar\varphi ) |$ reads
\begin{equation}
I_c (\alpha )= \bar I (\pi -2\alpha )= I_c J_0(\alpha )  \cos \alpha .
\label{crcur}
\end{equation}
For $\alpha \gtrsim 1.345$ and $0< \bar\varphi < \pi$ the absolute maximum of $|\bar I(\bar\varphi ) |$ is reached at negative current values (cf. Fig. \ref{Fig6}) and the system switches to the $\pi$-junction state which persists up to $\alpha \simeq 2.9$ where $I_c(\alpha )$ achieves its second local minimum (see Fig. \ref{Fig8}). In this regime for $1.7 \lesssim \alpha \lesssim 2.7$ the part of CPR with $d\bar I/d \bar \varphi >0$ is well approximated by the dependence $\bar I /I_c =-a\cos (\varphi /2)$ with $a \simeq 0.46$. At $\alpha \gtrsim 2.9$ CPR again describes a 0-junction state (cf. Fig. \ref{Fig7}).

In the limit of large values of $\alpha \gg 1$ CPR can be approximately described by keeping only the first term ($m=1$) in the series \eqref{CPRac} which takes the form
\begin{equation}
\bar I (\bar \varphi ) \simeq \frac {8I_c}{3\pi^{3/2}\sqrt{\alpha}}\sin \bar \varphi \cos \left(2\alpha -\frac{\pi}{4} \right).
\label{la}
\end{equation}
This dependence works reasonably well except in the immediate vicinity of the points $\alpha = 3\pi /8 +\pi p/2$ ($p=0,\pm 1, \pm 2, ...$) where terms with $m>1$ in Eq. \eqref{CPRac} need to be retained as well.

The dependence of the critical current on $\alpha$ at $T \to 0$ is displayed in Fig. \ref{Fig8}. We observe that -- in contrast to the case of Josephson tunnel junctions \cite{BP} -- $I_c (\alpha )$ does not vanish at any finite value of $\alpha$, it demonstrates oscillations which decay
with increasing $\alpha$ as $I_c (\alpha) \propto 1/\sqrt{\alpha }$. Local minima of $I_c (\alpha )$ correspond to consecutive transitions between 0- and $\pi$-junction states. Note that such transitions take place at non-zero values of the critical current. A similar feature was recently predicted \cite{KDZ20,KZ21} for $X$-junctions driven out of equilibrium by applying a temperature gradient. Hence, this feature is likely to be generic for junctions with non-sinusoidal CPR.  We also note that the behavior of the critical current $I_c (\alpha )$ displayed in Fig. \ref{Fig8} is fully consistent with that reported previously \cite{Bergeret}.
In order to extend our analysis beyond the adiabatic limit \eqref{adco} it is necessary to include resonances associated with excitation of the system from the lower Andreev state to the upper one by absorption of $l$ photons ($l=1,2,...$) with frequency $\omega <2\Delta$. This resonant process boosts the (negative) contribution to the supercurrent generated by the upper Andreev level and, hence, causes narrow dips on CPR \cite{Bergeret} at the phase values $\bar \varphi = \varphi_l$ ($l=1,2,...$) obeying the condition $l \omega = |E^A_{+} ( \varphi_l)- E^A_{-} (\varphi_l)|$. Obviously, these dips are not captured within our simple adiabatic approximation. They, however, do not lead to any qualitative modifications of the dependence of the critical current on $\alpha$ in Fig. \ref{Fig8}. At even higher frequencies $\omega \gtrsim 2\Delta$ the critical current gets further decreased due to pairbreaking effects.

Finally, we stress that the above results apply in the interesting for us low temperature limit $T \ll \Delta$. With increasing temperature CPR quickly approaches the sinusoidal form and the adiabatic approximation yields vanishing critical current at the values of $\alpha$ where the Bessel function $J_0 (2\alpha)$ has its zeros. On the other hand, at not too low temperatures non-equilibrium effects caused by external radiation may also stimulate the supercurrent \cite{Zaikin83}, i.e. under certain conditions the critical current in SNS junctions may actually increase in the presence of an ac signal \cite{Zaikin83,Bergeret}.

\section{Shapiro steps}
Let us now assume that our SNS junction is biased by an external voltage $V(t)$ which contains both constant in time and ac parts, i.e.
\begin{equation}
V(t)=V+V_{\rm ac}\cos (\omega t+\vartheta),
\label{vt}
\end{equation}
where $\vartheta$ is an arbitrary phase. Accordingly, for the Josephson phase we now get
\begin{equation}
\varphi (t)= 2e\int^tdt'V(t')=2eVt+2\alpha\sin(\omega t+\vartheta).
\label{acv}
\end{equation}
As before, here the parameter $\alpha$ \eqref{phase} effectively controls the strength of microwave radiation effects.

The overall effect of external microwave radiation on the $I-V$ curve in fully transparent SNS junctions was addressed in Ref. \onlinecite{Uwe2}. The physics behind this effect is transparent: While suffering MAR inside the junction quasiparticles and holes may now also absorb and emit photons with frequency $\omega$. As a result of such processes, the quasiparticle energy as well as its distribution function may change during the MAR cycle which, in turn, yields three major consequences \cite{Uwe2}: (i) Reduction of zero bias conductance and excess current (particularly pronounced at smaller bias voltages $eV \lesssim 2\Delta$), (ii) substantial modification of the subharmonic gap structure on the $I-V$ curve and (iii) appearance of Shapiro steps at microwave radiation frequencies (\ref{Shst}).

Despite some previous efforts \cite{AB2,Uwe2,Cuevas} a detailed microscopic theory of Shapiro steps in ballistic SNS junctions is not yet completed. Below we will employ our formalism in order to achieve some progress in this direction.

\subsection{Small microwave signals}
We first assume that this parameter is small, $\alpha\ll 1$, and evaluate the effect of external radiation on the $I-V$ curve perturbatively in $\alpha$. The linear in $\alpha$ corrections $\delta \hat X^{R,A,K}$ to (already evaluated in the previous section) matrices $\hat X^{R,A,K}$ read
\begin{eqnarray}
\delta \hat X^R=-\hat X^R\delta \hat a^R \hat X^R,\;\delta \hat X^A=\hat X^A\delta \hat a^A \hat X^A,\\
\delta \hat X^K=-\hat X^R\delta \hat a^K \hat X^A-\hat X^R\delta \hat a^R \hat X^K+\hat X^K\delta \hat a^A \hat X^A.
\end{eqnarray}
We also expand the matrices $\hat a^{R,A,K}$ and  $\hat b^{R,A,K}$ (cf. Eqs. \eqref{adef} and \eqref{bexp}) up to the linear in $\alpha$ terms
\begin{widetext}
\begin{eqnarray}
\delta \hat a^{R}(t,t')=-\frac{\alpha}{2} \left(\begin{array}{cc} g^{R}(t,t')\sin\frac{eV(t-t')}{2}\left(\sin(\omega t+\vartheta)-\sin (\omega t'+\vartheta)\right) &  f^{R}(t,t')\sin\frac{eV(t+t')}{2}\left(\sin(\omega t+\vartheta)+\sin (\omega t'+\vartheta)\right)
\\ f^{R}(t,t')\sin\frac{eV(t+t')}{2}\left(\sin(\omega t+\vartheta)+\sin (\omega t'+\vartheta)\right) & g^{R}(t,t')\sin\frac{eV(t-t')}{2}\left(\sin(\omega t+\vartheta)-\sin (\omega t'+\vartheta)\right)
 \end{array} \right),\nonumber
 \end{eqnarray}
\begin{eqnarray}
\delta \hat b^{R}=\frac{\alpha}{2}\left(\begin{array}{cc} g^{R}(t,t')\cos\frac{eV(t-t')}{2}\left(\sin(\omega t+\vartheta)-\sin (\omega t'+\vartheta)\right) &  f^{R}(t,t')\cos\frac{eV(t+t')}{2}\left(\sin(\omega t+\vartheta)+\sin (\omega t'+\vartheta)\right)
\\ f^{R}(t,t')\cos\frac{eV(t+t')}{2}\left(\sin(\omega t+\vartheta)+\sin (\omega t'+\vartheta)\right) & g^{R}(t,t')\cos\frac{eV(t-t')}{2}\left(\sin(\omega t+\vartheta)-\sin (\omega t'+\vartheta)\right)
 \end{array} \right)\nonumber
\end{eqnarray}
\end{widetext}
and similarly for $a^{A,K}$ and $b^{A,K}$.  Combining all these expressions with Eq. (\ref{abstcu}) we arrive at the linear in $\alpha$
correction to the $I-V$ curve of our voltage-biased SNS junction due to the presence of external radiation. It is easy to observe that
this correction differs from zero only under the condition $\omega =k\omega_J$ with $k=1,2,...$ giving rise to integer $\vartheta$-dependent Shapiro steps $\delta I_k (\vartheta)$.

Explicit analytic expressions for $\delta I_k$ are specified in Appendix B, cf. Eq. (\ref{genexp2}).  They take the form
\begin{equation}
\delta I_k(\vartheta)=a_k \cos\vartheta + b_k \sin\vartheta,
\end{equation}
implying that the overall magnitude of the $k$-th Shapiro step is $\delta I_k=2\sqrt{a_k^2+b_k^2}$. Employing Eq. (\ref{genexp2})
one can evaluate $\delta I_k$ numerically for different values of radiation frequency $\omega$, temperature $T$ and effective inelastic relaxation rate $\gamma$.

\begin{figure}
\includegraphics[width=8cm]{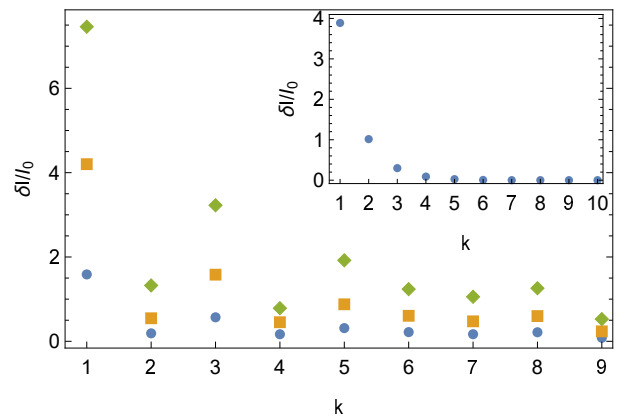}
\caption{The magnitude of integer Shapiro steps $\delta I_k$ (normalized by $I_0= \alpha \Delta /(2eR_N)$) evaluated from Eq. (\ref{genexp2}) for $\omega /\Delta =0.5$,  $\gamma/\Delta =0.01$ and different values of $T/\Delta=0.1$ (diamonds), 1 (squares) and 3 (circles).  Inset:
The same for  $\omega /\Delta =0.5$, $T/\Delta=0.1$ and $\gamma/\Delta =0.2$.}
\label{combfig1}
\end{figure}

\begin{figure}
\includegraphics[width=8cm]{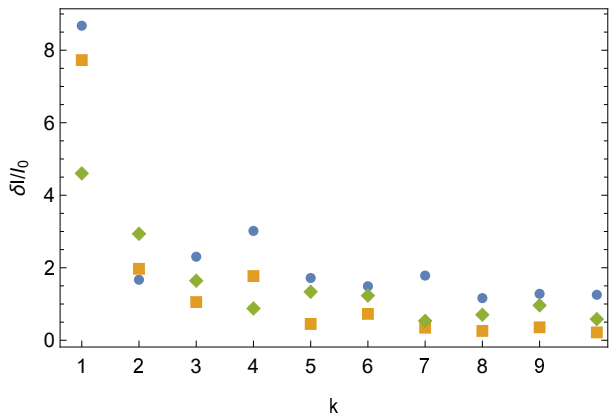}
\caption{ The same as in Fig. \ref{combfig1} for $T/\Delta=0.1$,  $\gamma/\Delta =0.01$ and different values of $\omega/\Delta =0.1$ (squares),  0.3 (diamonds) and 1.2 (circles). The steps indicated by diamonds are shifted by 0.4 upwards to avoid overlapping. }
\label{combfig2}
\end{figure}

Figure  \ref{combfig1} demonstrates the dependence of integer Shapiro steps $\delta I_k$ on $k$ at three different values of $T$ and the external radiation frequency $\omega=0.5\Delta$ in the limit of very weak inelastic relaxation $\gamma=0.01 \Delta$ in which case the phase coherence remains well preserved. We observe that $\delta I_k$ demonstrates decaying oscillations with increasing $k$. In addition, there is an overall decrease of $\delta I_k$ with increasing $T$ at any given $k$ due to temperature smearing. Note that such oscillations disappear completely even at low values of $T$ as soon as inelastic relaxation becomes more pronounced. This effect is illustrated in the inset of Fig.  \ref{combfig1} demonstrating $\delta I_k$ decaying monotonously with increasing $k$.

Figure \ref{combfig2} displays the dependence of the magnitude of Shapiro steps on $k$ at different values of $\omega$ for $\gamma=0.01 \Delta$ and $T=0.1\Delta$. We again observe decaying oscillations of $\delta I_k$ with increasing $k$.

\subsection{Beyond perturbation theory}
The above perturbative in $\alpha$ procedure is sufficient to evaluate all integer Shapiro steps with $\omega=k\omega_J$, whereas
in order to recover fractional Shapiro steps (\ref{Shst}) with $n>1$ corresponding to multi-photon processes it is necessary to proceed to
higher orders in $\alpha$. Accordingly, for $\alpha \ll 1$ such fractional steps remain parametrically smaller than the integer ones and,
hence, can be neglected in this limit. Note, however, that for small values of the bias voltage $V$ the parameter $\alpha =eV_{\rm ac}/\omega \sim V_{\rm ac}/V$ is in general not at all small, thus making any perturbation theory in $\alpha$ insufficient. The task at hand is to go beyond such perturbation theory and evaluate both integer and fractional Shapiro steps for arbitrary values of $\alpha$.

In order to accomplish this goal one again has to compute the inverse matrices $\hat X^{R,A}= (\hat a^{R,A})^{-1}$ and substitute the result into
Eq. \eqref{abstcu} for the current. The procedure in the presence of an ac signal \eqref{vt} is outlined in Appendix C. Here, however, we take a somewhat different route. We will first give a simple estimate for fractional Shapiro steps which turns out to be sufficient for $\alpha \lesssim 1$. Then we will specifically address the interesting for us low bias regime in which case one can develop a more accurate calculation that will be applicable for all values of $\alpha$.

Our simple estimate for the magnitude of (fractional) Shapiro steps $\delta I_{\frac1n}$ which occur at $\omega=\omega_J/n$ is outlined in Appendix D. It yields
\begin{equation}
\frac{\delta I_{\frac1n}}{I_c}\sim \left\{\begin{array}{l} J^2_{n/2}\left( \alpha/2\right)\;{\rm for\, even}\; n,\\ |J_{(n+1)/2}\left( \alpha/2\right)J_{(n-1)/2}\left( \alpha/2\right)|\;{\rm for\, odd}\; n, \end{array} \right.
\label{estim}
\end{equation}
where, as above, $J_n(x)$ are Bessel functions.

All these Shapiro steps correspond to $k=1$. The magnitude of other fractional Shapiro steps $\delta I_{\frac{k}{n}}$ decays with growing $k$ similarly to integer Shapiro steps considered above.

It is obvious from the above estimate that for $\alpha \sim 1$ the magnitude of fractional Shapiro steps is in general of the same order as that for integer Shapiro steps. Furthermore, the former can easily exceed the latter for some values of $\alpha$. For instance, the ratio between the magnitudes of Shapiro steps $\delta I_1$ and $\delta I_{\frac12}$ which occur respectively at $\omega=\omega_J$ and $\omega=\omega_J/2$
reads
\begin{equation}
\frac{\delta I_{1}}{\delta I_{\frac12}} \sim \frac{|J_0(\alpha )|}{|J_1(\alpha)|}.
\end{equation}
Clearly, for $\alpha \ll 1$ the integer Shapiro step always dominates $\delta I_{1} \gg \delta I_{\frac12}$, whereas for $\alpha \sim 1$ we already have $\delta I_{1} \sim \delta I_{\frac12}$ or possibly even $\delta I_{1} \ll \delta I_{\frac12}$ since Bessel functions of different order have different zeroes. Hence, for small enough voltages $V$ the fractional ("$4\pi$") Shapiro step can dominate of the integer ("$2\pi$") one even though the so-called "$4\pi$-Josephson effect" is totally absent. In fact, this observation should not be taken as any surprise because
it is well known that in the case of tunnel junctions one has \cite{Tinkh,BP,Likh} $\delta I_{\frac1n} \propto J_n(2\alpha)$.

\begin{figure}
\includegraphics[width=8cm]{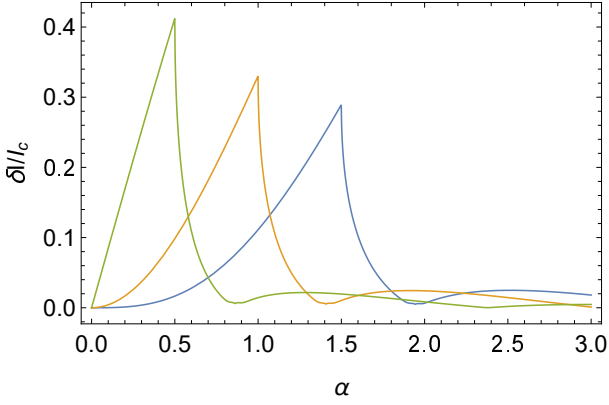}
\caption{Shapiro steps $\delta I_{1}(\alpha)$ (green), $\delta I_{\frac12}(\alpha)$ (orange) and $\delta I_{\frac13}(\alpha)$ (blue)
derived from Eq. \eqref{nev} respectively at $\omega=\omega_J$, $\omega=\omega_J/2$ and $\omega=\omega_J/3$. For $\alpha>0$ these three curves reach their first minima respectively at $\alpha\simeq$ 0.862,\, 1.407 and 1.943 where they take very small but positive values min$(\delta I_{1})\simeq 0.006$,  min$(\delta I_{\frac12}) \simeq 0.0053$ and min$(\delta I_{\frac13}) \simeq 0.005$. Note that the above numbers as well as the sharp peak feature observed in all these curves may to a certain extent depend on the approximation employed here.}
\label{combfig3}
\end{figure}

At sufficiently small values of the applied voltage $eV \ll \Delta$ it is also possible to evaluate the magnitude of fractional Shapiro steps more accurately by extending Eq. \eqref{smallVIphi} to the time-dependent bias voltage in the form \eqref{vt}. The accuracy of this approximation will be discussed below in the next section. Substituting the phase $\varphi (t)$ \eqref{acv} into Eq. \eqref{smallVIphi} and averaging the resulting current $I(t)$ over time
we obtain
\begin{eqnarray}
\nonumber
\bar I=\frac{V}{R_N}+I_c\left\langle\left|\sin \left[ z +\vartheta +\alpha\sin\left(\frac{\omega}{eV} z \right)\right]\right| \right.\\
\left.\times {\rm sgn}\left[1+\frac{\alpha\omega}{eV}\cos\left(\frac{\omega}{eV} z \right)\right]\right\rangle_z
\label{nev}
\end{eqnarray}
Here we are interested in the $\vartheta$-dependent terms emerging from Eq. \eqref{nev} provided $\omega$ and $eV$ are commensurate. In this case the current periodically depends on the phase $\vartheta$  reaching both maximum and minimum values $I_{\rm max}$ and $I_{\rm min}$ within the period $\pi$. The magnitude of a Shapiro step is then defined as $\delta I=I_{\rm max}-I_{\rm min}$.
Three different Shapiro steps  $\delta I_{1}$, $\delta I_{\frac12}$ and $\delta I_{\frac13}$ corresponding respectively to $\omega=2eV$, $\omega=eV$ and $\omega=2eV/3$ are displayed in Fig. \ref{combfig3} as functions of the parameter $\alpha$.

In accordance with our perturbative results we observe that at small values of $\alpha$ the principal Shapiro step  $\delta I_{1} \propto \alpha$ dominates over the fractional ones $\delta I_{\frac1n} \propto \alpha^n$. On the other hand, the peaks of all three curves  at $\alpha \sim 1$ are comparable in magnitude and each of the Shapiro steps can be significantly larger than the other ones due to different positions of these peaks. For instance,  at low enough voltages $V$ within the interval $0.6 \lesssim \alpha \lesssim 1.3$ the fractional Shapiro step $\delta I_{\frac12}$ dominates over $\delta I_{1}$ which almost vanishes in the middle part of this interval. On the other hand, at higher voltages the condition $\omega =2eV$ is fulfilled al lower values of $\alpha$ where the integer step $\delta I_{1}(\alpha)$ already becomes significant. This behavior is qualitatively consistent with the one observed in experiments  \cite{M16,topmat,Gre} but, of course, it by no means can serve as an evidence for $4\pi$-periodic CPR in our junctions.

Comparing our simple estimate \eqref{estim} with more accurate results following from Eq. \eqref{nev}  we conclude that -- in the agreement with our expectations (see Appendix D) -- the former works reasonably well for $\alpha\lesssim 1$ whereas for larger values of $\alpha$ the fractional Shapiro steps are better approximated, e.g., by the formula $\delta I_{\frac1n} \sim I_c J_{n/2}^2(3\alpha)$, cf. Fig. \ref{combfig3}.

\section{Current-biased regime}
Let us now consider a somewhat different physical situation assuming that our SNS junction is biased by an external
current in the form
\begin{equation}
I(t)=I+I_{\rm ac}\cos (\omega t + \vartheta).
\label{cb}
\end{equation}
Such current biased regime appears to be realized in a number of recent experiments  \cite{M16,topmat,Gre}. It is obvious that in this case the voltage across the junction cannot anymore be constant and should vary in time $V=V(t)$ even in the absence of an ac signal $I_{\rm ac}\to 0$.

Within the adiabatic approximation, i.e. provided the voltage $V(t)$ changes in time slowly enough, it is still possible to employ the same CPR \eqref{smallVIphi} and \eqref{largeVIphi} initially derived for the time independent voltage bias. This approximation should work provided $V(t)$  does not change significantly during the whole MAR cycle, i.e. during the period of time
\begin{equation}
\delta t_{\rm mar}  \sim \frac{2\Delta}{eV(t)}\frac{d}{v_F}
\label{tMAR}
\end{equation}
it takes for a quasiparticle to cross the normal layer between two superconductors $\sim 2\Delta /(eV(t))$ times and escape into one of the superconducting terminals, see Fig. 2.

In fact, we have already made use of this approximation in the previous section evaluating the $I-V$ curve and Shapiro steps in the presence of  an ac voltage signal \eqref{vt} simply by substituting the Josephson phase in the form \eqref{acv} into CPR  \eqref{smallVIphi}.  In that case it suffices to require $\delta t_{\rm mar} \ll 1/\omega$ or, equivalently, $d \ll \xi_0eV(t)/\omega$.  Here, we will also assume that $V(t)$ changes slowly enough and again stick to the same adiabatic approximation which we verify in the end of our calculation. Within this approximation and provided $eV(t)$ remains much smaller than $\Delta$ the phase and voltage dynamics is controlled by  the following equation
\begin{equation}
I+I_{\rm ac}\cos(\omega t + \vartheta)=\dot \varphi /(2eR_N) +I_c|\sin (\varphi /2)|\sgn \dot \varphi,
\label{eqI}
\end{equation}
which describes the current balance in our SNS junction.

In the absence of an ac current component, i.e. in the limit $I_{\rm ac}\to 0$, the solution of this equation can be written in the form
\begin{equation}
F(\varphi ,a)=eI_cR_Nt, \quad a=I/I_c >1,
\label{sol}
\end{equation}
where
\begin{eqnarray}
&&F(\varphi ,a)=\frac{2}{\sqrt{a^2-1}}\left[ \arctan\left( \frac{a\tan \left(\frac{\varphi}{4}\right)-{\rm sgn}\left[\tan \left(\frac{\varphi}{4}\right)\right]}{\sqrt{a^2-1}}\right)\right.\nonumber\\
&&\left. +\pi\chi_1+\left(2 \arctan\frac{1}{\sqrt{a^2-1}}-\pi\right)
\chi_2\right], \label{F}
\end{eqnarray}
where we defined $\chi_n={\rm floor}\left[\varphi/(2\pi n)\right]$ ($n=1,2$) as integer parts of $\varphi/(2\pi n)$. This solution allows to determine the oscillating in time voltage across the junction $V(t)=\dot\varphi /2e$ which reads
\begin{equation}
V(t)=\frac{I_cR_N(a^2-1)}{a+\sin\left[\sqrt{a^2-1}\left|s\right|-\arctan(1/\sqrt{a^2-1}) \right]},
\label{Vt}
\end{equation}
where the parameter
\begin{eqnarray}
s= eI_cR_Nt+\frac{2}{\sqrt{a^2-1}}\arctan\frac{1}{\sqrt{a^2-1}}
\end{eqnarray}
is defined within the interval
\begin{eqnarray}
-\frac{\pi b}{\sqrt{a^2-1}}<s<\frac{\pi b}{\sqrt{a^2-1}}, \quad b=\frac{2}{\pi}\arctan\sqrt{\frac{a+1}{a-1}}.
\end{eqnarray}
The function $V(t)$ is continued periodically outside this interval with the period $2\pi b/(eI_cR_N\sqrt{a^2-1})$.

The average voltage $\bar V$ is determined either by averaging of Eq. \eqref{Vt} over this period or simply by combining the asymptotic form
$F(\varphi \gg 1,a)\simeq \varphi b/\sqrt{a^2-1}$ with Eq. \eqref{sol}. Either of these two ways yields the following $I-V$ curve for our SNS junction (cf. also \cite{AB2})
\begin{eqnarray}
\bar V= \frac{\pi}{4}I_cR_N\frac{\sqrt{(I/I_c)^2-1}}{\arctan\sqrt{\frac{I+I_c}{I-I_c}}}.
\label{IVcb}
\end{eqnarray}
Bearing in mind the relation $I_cR_N=\pi\Delta/e$ we note that the condition $e\bar{V} \ll \Delta$ implies that Eqs. \eqref{Vt} and \eqref{IVcb} remain valid as long as $I-I_c \ll I_c$. In this limit from
Eq. \eqref{IVcb} we obtain
\begin{eqnarray}
 \bar V\simeq \frac{\pi\Delta}{2e}\sqrt{(I/I_c)^2-1}, \quad I-I_c \ll I_c.
\end{eqnarray}
In addition, with the aid of Eqs. \eqref{tMAR} and \eqref{Vt} we conclude that our adiabatic approximation holds for sufficiently short SNS junctions with $d \ll \xi_0$.

Let us now include an ac current bias into our consideration. Provided the amplitude of an ac signal remains small as compared to the critical current $I_{\rm ac} \ll I_c$ one can solve Eq. \eqref{eqI} perturbatively in $I_{\rm ac}$. In the first order in this parameter we obtain
\begin{equation}
\delta \varphi (t) = 2eR_N V(t) \int^tdt' \frac{\delta I +I_{\rm ac}\cos (\omega t'+\vartheta )}{V(t')},
\label{dphi}
\end{equation}
where $V(t)$ is defined in \eqref{Vt} and $\delta I$ denotes the correction to the current flowing across the junction. The time derivative of the first order correction to the phase (\ref{dphi}) determines an extra voltage value $\delta V= \delta \dot \varphi \propto I_{\rm ac}$ generated by an ac current signal. It is necessary to verify that the time average $\langle \delta V\rangle_t=\delta\varphi (t)/t|_{t \to \infty}$ equals to zero \cite{AL,Th}. At all frequencies except for particular values $\omega =2ek\bar{V}$ this condition is justified for $\delta I=0$ in which case we are back to the $I-V$ curve \eqref{IVcb} derived in the absence of an ac signal. For $\omega=2ek\bar{V}$ we arrive at integer Shapiro steps in the form
\begin{eqnarray}
\label{cbsm}
\delta I_k(\vartheta) =-\frac{I_cI_{\rm ac} b \sin(\pi b)\cos \vartheta}{\pi I \left(b^2-k^2\right)}
\left[ 1+\frac{\sqrt{a^2-1}}{\pi a^2 b}\right]^{-1}.
\label{shst}
\end{eqnarray}
In the interesting for us limit  $I-I_c \ll I_c$ Eq. \eqref{shst} yields $\delta I_{k=1}\sim I_{\rm ac}$ and $\delta I_{k\geq2}\sim I_{\rm ac}\sqrt{a-1}/k^2$, i.e. the amplitudes of all Shapiro steps with $k\geq 2$ contain extra small factor $\sqrt{a-1} \ll 1$ and, on top of that, decay quickly with increasing $k$.

As before, fractional Shapiro steps emerge only in the higher orders in the ac signal and are mostly pronounced for sufficiently large values of  $I_{\rm ac}$. In the latter limit Eq. \eqref{eqI} can easily be resolved perturbatively in $I_c$. In the lowest order one can neglect the last term in Eq. \eqref{eqI} and then easily integrate this equation with the result $\varphi (t) = 2eIR_Nt +2\alpha \sin (\omega t +\vartheta)$. Substituting this expression for $\varphi (t)$ into the last term in Eq. \eqref{eqI} and averaging the result over time we again arrive at Eq. (\ref{nev}) with
\begin{equation}
\bar I \to I, \quad V \to \bar V, \quad \alpha=\frac{eI_{\rm ac} R_N}{\omega}.
\end{equation}
As we already know, this equation describes fractional Shapiro steps  $\delta I_{\frac1n}$ which occur at frequencies $\omega=2e\bar V/n$. For small $\alpha$ one again has $\delta I_{\frac1n} \propto \alpha^n$, whereas for bigger values of $\alpha$ the fractional Shapiro steps $\delta I_{\frac1n}(\alpha)$ are displayed in Fig. \ref{combfig3}. Note that the interesting for us parameter range $\alpha \sim 1$ (or, equivalently, $I_{\rm ac} \sim I_c$) is now close to the border of applicability of our approach and, hence, for such values of $\alpha$ one could in principle expect some corrections to Eq. \eqref{nev} which, however, cannot alter any of our conclusions. In particular, we may conclude that fractional Shapiro steps can dominate over integer ones at sufficiently small frequencies $\omega$ of an external ac signal (i.e. for $\alpha \gtrsim 1$) also in the current-biased regime.

\section{Discussion}
In this work we constructed a microscopic theory describing Shapiro steps in topologically trivial ballistic SNS junctions subject to external microwave radiation of arbitrary intensity. Our analysis was mainly focused on the most interesting limit of subgap voltages and temperatures revealing a number of features similar to those observed in recent experiments \cite{M16,topmat,Gre}
with superconducting junctions involving topological insulators. The main such features are: i) the presence of non-vanishing "$n=0$ Shapiro step", i.e. the critical current $I_c (\alpha )$ that does not vanish at any finite $\alpha$, ii) the presence of Shapiro steps on the $I-V$ curve at all voltages and frequencies obeying the condition \eqref{Shst}, including the "fractional" one with $\omega =\omega_J/2$ sometimes interpreted as an evidence for the so-called $4\pi$-periodic Josephson effect and iii) the possibility for "missing" Shapiro step with $\omega =\omega_J$ (along with other Shapiro steps) at certain radiation intensities and frequencies.

The feature i) is a direct consequence of a non-sinusoidal CPR inherent to all types of highly transparent superconducting weak links at low temperatures $T \ll \Delta$. Hence, it is by no means surprising that this feature has been observed in experiments  \cite{Gre} where
junctions with high quality interfaces have been employed.

The feature ii) has to do with multi-photon processes. They play little role at low intensities of microwave radiation (i.e. for $\alpha \ll 1$)
in which case integer Shapiro steps  $\delta I_{k}(\alpha)$ (with $k \geq 1$) are mainly pronounced, see Figs.  \ref{combfig1} and
\ref{combfig2}. However, at higher ac signals or smaller microwave frequencies  fractional Shapiro steps   $\delta I_{\frac1n}(\alpha)$ (with $n \geq 2$) gain importance reaching almost the same magnitude as integer ones for $\alpha \gtrsim 1$, cf., e.g., Fig. \ref{combfig3}.

Finally, the feature iii), i.e. "missing" Shapiro steps, can be explained with the aid of our results displayed in Fig. \ref{combfig3}. Obviously, at around $\alpha \approx 1$  the step $\delta I_{1}(\alpha)$ almost vanishes being much smaller than, e.g., $\delta I_{\frac12}(\alpha)$ and $\delta I_{\frac13}(\alpha)$. Keeping the amplitude of an ac signal unchanged and increasing the radiation frequency $\omega$ one decreases $\alpha$ eventually reaching the regime where $\delta I_{1}(\alpha)$ becomes of order or even bigger than fractional Shapiro steps. Then in the experiment one would observe either "missing" Shapiro step $\delta I_{1}(\alpha)$ or, to the contrary, fully restored one respectively at the lower and higher radiation frequencies $\omega$. This is exactly what has been observed in the experiments \cite{M16,topmat,Gre}.

Thus, our analysis demonstrates that all experimentally detected features i), ii) and iii) can be reproduced within a microscopic model describing topologically trivial SNS junctions with $2\pi$-periodic CPR, cf., e.g., the results presented in Figs. \ref{Fig8} and \ref{combfig3} of this work and those in Fig. 2 of Ref. \onlinecite{Gre}. None of these features actually requires $4\pi$-periodic CPR which never pops up in our calculation. In addition, we point out that missing integer Shapiro steps at $\omega =\omega_J$ -- along with well pronounced fractional ones -- were  recently observed in topologically trivial Josephson junctions based on InAs quantum wells \cite{MS}. It is also well known that Shapiro steps with odd $n=1,3,...$ can be significantly reduced (as compared to ones with even $n=2,4,...$) or even vanish completely due to size effects \cite{FN1}. Hence, caution is needed while unambiguously interpreting experimental results for superconducting weak links in terms of Majorana physics.

We believe that all the results obtained here can be directly applied also to superconducting junctions hosting Majorana-like bound states \eqref{EM} in the case of preserved time-reversal symmetry $\delta \approx \Delta$. No $4\pi$-periodic CPR is expected for any voltage bias $V$ in that case. The situation with $\delta < \Delta$ may still deserve a separate microscopic analysis.

Note, however, that also in the latter case the treatment of ac Josephson effect exploiting Eq. \eqref{dEdf} generally requires a more rigorous justification.  In the limit of short junctions this equation is verified only in equilibrium in which case it can also be extended to more complicated situations, e.g., to hybrid superconducting structures involving triplet pairing \cite{GKZ,KGZ}. The situation out of equilibrium could be more tricky since, e.g., the effect of MAR needs to be included into consideration. At not too small junction transmissions MAR sets in already at low voltages and, as we have demonstrated, becomes an essential ingredient of the whole physical picture. Multiple Andreev reflection also plays an important role in experiments \cite{M16,topmat,Gre} with junctions involving topological insulators \cite{FN2} and, hence, it needs to be properly accounted for in any theoretical analysis of ac Josephson effect in such structures.

\vspace{0.5cm}

\centerline{\bf Acknowledgements}

We acknowledge useful discussions with M.S. Kalenkov and A.G. Semenov.

\vspace{0.5cm}

\appendix

\section{Matrix inversion}
To begin with, let us introduce convenient notations for time dependent quasiclassical propagators employed in our calculation. These propagators in general depend on two time variables. They can be expressed in terms of the following expansion over the voltage harmonics
\begin{equation}
x(t,t')=\sum_{n=-\infty}^\infty \int\frac{d\epsilon}{2\pi} x(\epsilon,n) e^{-i\epsilon(t-t')} e^{-i ne V(t+t')/2},\label{feq}
\end{equation}
which defines $x(\epsilon,n)$. We will also make use of the notation $z(t,n)$ which essentially means that
$$z(t,t')=z(t-t')\exp(-i ne V(t+t')/2).$$

Below we will employ the following property: Provided  $z(t,t')$ can be represented as a convolution
$$z(t,t')=f(t,m)\circ g(t,n),$$ then $z(\epsilon, m+n)$ has only $m+n$ components with
\begin{equation}
z(\epsilon,m+n)=f\left(\epsilon+\frac{neV}{2},m\right)g\left(\epsilon-\frac{meV}{2},n\right).\label{crule}
\end{equation}

In order to proceed let us express the matrix $\hat X^R(\epsilon,n)$ in the form
\begin{equation}
\hat X^R(\epsilon,n)=\left(\begin{array}{cc} \beta^R(\epsilon,n) & \gamma^R(\epsilon,n)\\ \gamma^R(\epsilon,n)& \beta^R(\epsilon,n) \end{array} \right).
\end{equation}
Note that $\beta^R(\epsilon,n)$ differs from zero for even $n$, while $\gamma^R(\epsilon,n)$ is non-zero for odd $n$. Then the matrix equation
\begin{equation}
(\hat a^R\circ \hat X^R)(\epsilon,n)=\delta_{n,0}
\end{equation}
can be identically rewritten in terms of the following equations
\begin{widetext}
\begin{eqnarray}
&& \frac{1}{2}\left[g^R\left(\epsilon+(n+1)\frac{eV}{2} \right) + g^R\left(\epsilon+(n-1)\frac{eV}{2} \right)\right] \beta^R(\epsilon,n)+\label{e1}\\ &&\frac{1}{2} f^R\left(\epsilon+(n-1)\frac{eV}{2} \right)\gamma^R\left(\epsilon-\frac{eV}{2}, n-1\right) +\frac{1}{2} f^R\left(\epsilon+(n+1)\frac{eV}{2} \right)\gamma^R\left(\epsilon+\frac{eV}{2}, n+1\right)=
\delta_{n,0}\nonumber
\end{eqnarray}
and
\begin{eqnarray}
&& \frac{1}{2}\left[g^R\left(\epsilon+(n+1)\frac{eV}{2} \right) + g^R\left(\epsilon+(n-1)\frac{eV}{2} \right)\right] \gamma^R(\epsilon,n)+\label{e2}\\ &&\frac{1}{2} f^R\left(\epsilon+(n-1)\frac{eV}{2} \right)\beta^R\left(\epsilon-\frac{eV}{2}, n-1\right) +\frac{1}{2} f^R\left(\epsilon+(n+1)\frac{eV}{2} \right)\beta^R\left(\epsilon+\frac{eV}{2}, n+1\right)=
0.\nonumber
\end{eqnarray}
It is convenient to introduce the variables $\zeta^R(\epsilon,n)$ and  $\tilde\zeta^R (\epsilon,n)$ which read
\begin{equation}
\zeta^R(\epsilon,n)=\left\{ \begin{array}{c} \beta^R(\epsilon,n)\quad {\rm for\; even}\; n,\\ \gamma^R(\epsilon,n)\quad {\rm for\; odd}\; n,\end{array}\right.\quad \tilde\zeta^R (\epsilon,n)= \zeta^R \left(\epsilon+\frac{neV}{2},n\right).
\label{sharg}
\end{equation}
Then Eqs. (\ref{e1}), (\ref{e2}) can be rewritten as
\begin{eqnarray}
&& \frac{1}{2}\left[g^R\left(\epsilon+(2n+1)\frac{eV}{2} \right) + g^R\left(\epsilon+(2n-1)\frac{eV}{2} \right)\right] \tilde\zeta^R (\epsilon,n) + \label{treq}\\\ && \frac{1}{2} f^R\left(\epsilon+(2n-1)\frac{eV}{2} \right)\tilde\zeta^R (\epsilon,n-1)+ \frac{1}{2} f^R\left(\epsilon+(2n+1)\frac{eV}{2} \right)\tilde\zeta^R (\epsilon,n+1)=\delta_{n,0}.
\nonumber
\end{eqnarray}

This equation has essentially the structure $\hat T\hat\zeta^R=\delta_{n,0}$, where $\hat\zeta^R$ is the column composed of $\tilde\zeta^R(\epsilon,n)$ and $\hat T$ is the symmetric tridiagonal matrix with the elements
\begin{eqnarray}
&& T_{n,n}=\frac{1}{2} \left[g^R\left(\epsilon+\left(n+\frac{1}{2}\right)eV\right)  + g^R\left(\epsilon+\left(n-\frac{1}{2}\right)eV\right)\right],\label{trim}\\
&& T_{n,n+1}=T_{n+1,n}=\frac{1}{2}f^R\left(\epsilon+\left(n+\frac{1}{2}\right)eV\right).\nonumber
\end{eqnarray}

The equations allowing to determine the matrix $\hat X^A$ are derived in a similar manner. We introduce
\begin{equation}
\hat X^A(\epsilon,n)=\left(\begin{array}{cc} \beta^A(\epsilon,n) & \gamma^A(\epsilon,n)\\ \gamma^A(\epsilon,n)& \beta^A(\epsilon,n) \end{array} \right),\quad \zeta^A(\epsilon,n)=\left\{ \begin{array}{c} \beta^A(\epsilon,n)\quad {\rm for\; even}\; n,\\ \gamma^A(\epsilon,n)\quad {\rm for\; odd}\; n,\end{array}\right. \quad \tilde\zeta^A(\epsilon,n)=\zeta^A\left(\epsilon- \frac{neV}{2},n\right).
\end{equation}
Making use of Eq. (\ref{arr}) we can rewrite the equation $-(\hat X^A\circ \hat a^A)(\epsilon,n)=\delta_{n,0}$ in the form
\begin{eqnarray}
&& \frac{1}{2}\tilde\zeta^A(\epsilon,n)\left[g^R\left(\epsilon-\left(n-\frac{1}{2}\right)eV\right)  + g^R\left(\epsilon-\left(n+\frac{1}{2}\right)eV\right)\right]^*+\label{aeq}\\&&
\frac{1}{2}\tilde\zeta^A(\epsilon,n-1)\left[ f^R\left(\epsilon-\left(n-\frac{1}{2}\right)eV\right)\right]^*
+\frac{1}{2}\tilde\zeta^A(\epsilon,n+1)\left[ f^R\left(\epsilon-\left(n+\frac{1}{2}\right)eV\right)\right]^*=\delta_{n,0}.\nonumber
\end{eqnarray}
Equation (\ref{aeq}) can again be expressed as $\hat T\hat\zeta^A=\delta_{n,0}$ with
\begin{eqnarray}
&& T_{n,n}=\frac{1}{2} \left[g^R\left(\epsilon-\left(n+\frac{1}{2}\right)eV\right)  + g^R\left(\epsilon-\left(n-\frac{1}{2}\right)eV\right)\right]^*\\
&& T_{n,n+1}=T_{n+1,n}=\frac{1}{2}\left[ f^R\left(\epsilon-\left(n+\frac{1}{2}\right)eV\right)\right]^*.\nonumber
\end{eqnarray}

The inversion procedure for tridiagonal and block tridiagonal matrices is outlined in Ref. \onlinecite{M}. Employing the theorem 2.3 from \cite{M} one can explicitly evaluate the inverse of tridiagonal matrix $\hat T$, defined by Eq. (\ref{trim}). To be more specific, let us choose $\hat T_{nk}$  with $-N\le n,k\le N$, i.e. we are now dealing with a $(2N+1)\times(2N+1)$ square matrix where $N$ is large. Then, employing the second Eq. (\ref{prop}), we set
\begin{equation}
g^R\left(\epsilon+eV\left(-N-\frac{1}{2}\right)\right)=g^R\left(\epsilon+eV\left(N+\frac{1}{2}\right)\right)=1.
\end{equation}
All the other functions involved in the matrix elements are defined in Eqs. (\ref{trim}). Making use of the theorem 2.3 from \cite{M} together with the first Eq. (\ref{prop}), we get
\begin{eqnarray}
&& \left( \hat T^{-1}\right)_{0,0}=1,\quad \left( \hat T^{-1} \right)_{n,0}=\prod_{n\le k\le -1} \left(-\frac{f^R\left(\epsilon+eV\left(k+\frac{1}{2}\right)\right)}{1+g^R\left(\epsilon+eV\left(k+\frac{1}{2}
\right)\right)} \right),\; {\rm if}\; n<0 \nonumber \\
&& \left( \hat T^{-1} \right)_{n,0}=\prod_{0\le k\le n-1} \left(-\frac{f^R\left(\epsilon+eV\left(k+\frac{1}{2}\right)\right)}{1+g^R\left(\epsilon+eV\left(k+\frac{1}{2}
\right)\right)} \right),\; {\rm if}\; n>0.\label{mel}
\end{eqnarray}
Introducing the notation \eqref{are}, employing the multiplicative structure of Eqs. (\ref{mel}) together with
the relationship
\begin{eqnarray}
&& \left(g^R(\epsilon+neV)-g^R(\epsilon+(n+1)eV)\right) \tilde\zeta^R\left(\epsilon+\frac{eV}{2},n\right)+f^R(\epsilon+neV)\tilde\zeta^R\left(\epsilon+\frac{eV}{2},n-1\right)
\nonumber\\&& -
f^R(\epsilon+(n+1)eV)\tilde\zeta^R\left(\epsilon+\frac{eV}{2},n+1\right)=-2\,{\rm sgn}\,n\;\tilde\zeta^R\left(\epsilon+\frac{eV}{2},n\right),
\end{eqnarray}
from Eq. (\ref{abstcu}) we  recover Eqs. \eqref{Fou}-\eqref{arharm}.

\section{Shapiro steps at small microwave signals}
Employing the perturbative in $\alpha$ procedure outlined in Sec. VIA and making use of Eqs. (\ref{prop}), (\ref{prop2}), we arrive at general expressions for Shapiro current steps $\delta I_k$ at $\omega=2keV$ which take the form
\begin{eqnarray}
&& \delta I_k=\, {\rm Re}\left\{\frac{\pi \alpha }{16eR_N}\sum_{n,p,q} \int\limits_{-\infty}^\infty \frac{d\epsilon}{2\pi}\left({\rm sign}\,p - {\rm sign}\,n \right)\tilde \zeta^R(\epsilon,\,q)\tilde\zeta^A(\epsilon,\,p) \tilde F(\epsilon,\,q) \right.\nonumber\\
&& \times \tilde\zeta^R\left(\epsilon-(p+n)eV,\,n \right) Z'\left(\epsilon-(p-q+n)\frac{eV}{2},\,p+q+n\right)\label{genexp2}
\\ && +\frac{\pi \alpha}{16eR_N}\sum_{n,m} \int\limits_{-\infty}^\infty \frac{d\epsilon}{2\pi}
\tilde \zeta^R(\epsilon-neV,\,n)Z' \left( \epsilon-\frac{(n-m)eV}{2},\,n+m\right) \widetilde{\widetilde{ F}}(\epsilon,m)\tilde\zeta^R\left(\epsilon,\,m \right)\nonumber
\\ && +\frac{\pi \alpha}{16eR_N}\sum_{n,m} \int\limits_{-\infty}^\infty \frac{d\epsilon}{2\pi}\,
{\rm sign}\,n\,\tilde \zeta^R\left(\epsilon-\frac{(n+m)eV}{2},\,n\right)Z'' \left( \epsilon,\,n+m\right) \tilde\zeta^A\left(\epsilon+\frac{(n+m)eV}{2},\,m \right)\nonumber
\\ && -\frac{\pi \alpha }{4eR_N} \sum_{n}\sum_{p+q=n} \int\limits_{-\infty}^\infty  \frac{d\epsilon}{2\pi} Y'\left(\epsilon- \frac{(p-q)eV}{2},\, n\right) \tilde\zeta^R\left(\epsilon,\,q\right) \tilde\zeta^A\left(\epsilon,\,p\right)\tilde F\left(\epsilon,\,q\right)\nonumber
\\ &&\left. + \frac{\pi \alpha }{4eR_N} \sum_n\int\limits_{-\infty}^\infty \frac{d\epsilon}{2\pi}\, Y''(\epsilon,n)
\tilde\zeta^R\left(\epsilon-\frac{neV}{2},\, n\right)\right\}.\nonumber
\end{eqnarray}
Here we defined
\begin{eqnarray}
&& Z'(\epsilon,\,n+m)=\left[ g^R\left(\epsilon+\left(k-\frac{1}{2}\right)eV \right)+ g^R\left(\epsilon+\left(\frac{1}{2}-k\right)eV \right)- g^R\left(\epsilon+\left(k+\frac{1}{2}\right)eV\right) \right.
\nonumber
\\&& \left. - g^R\left(\epsilon-\left(k+\frac{1}{2}\right)eV \right)\right]e^{i\vartheta}\delta\left(2k-n-m \right)+\left[ f^R\left( \epsilon+keV\right)+ f^R\left( \epsilon-keV\right)\right]e^{i\vartheta}\nonumber\\&& \times\left[ \delta\left(2k-1-n-m \right) -\delta\left(2k+1-n-m\right)  \right]\label{pz}
\end{eqnarray}
and
\begin{eqnarray}
&& Z''(\epsilon,\,n+m)=\left[ g^K\left(\epsilon+\left(k-\frac{1}{2}\right)eV \right)+ g^K\left(\epsilon+\left(\frac{1}{2}-k\right)eV \right)- g^K\left(\epsilon+\left(k+\frac{1}{2}\right)eV\right) \right.
\nonumber
\\&& \left. - g^K\left(\epsilon-\left(k+\frac{1}{2}\right)eV \right)\right]e^{i\vartheta}\delta\left(2k-n-m \right)+\left[ f^K\left( \epsilon+keV\right)+ f^K
\left( \epsilon-keV\right)\right]e^{i\vartheta}\nonumber\\&& \times\left[ \delta\left(2k-1-n-m \right) -\delta\left(2k+1-n-m\right)  \right].\label{ppz}
\end{eqnarray}
Similarly we defined
\begin{eqnarray}
&& Y'(\epsilon,n)=\left[ g^R\left(\epsilon+\left(k+\frac{1}{2}\right)eV  \right)+g^R\left(\epsilon+\left(k-\frac{1}{2}\right)eV\right) -g^R\left(\epsilon-\left(k+\frac{1}{2}\right)eV  \right) \right.\nonumber\\ &&\left. -g^R\left(\epsilon-\left(k-\frac{1}{2}\right)eV \right)\right]e^{i\vartheta}\delta(2k-n)+
\left[f^R(\epsilon+keV) +  f^R(\epsilon-keV) \right]e^{i\vartheta}\nonumber\\&&\times\left[ \delta(2k+1-n)+\delta(2k-1-n) \right]
\end{eqnarray}
and
\begin{eqnarray}
&& Y''(\epsilon,n)=\left[ g^K\left(\epsilon+\left(k+\frac{1}{2}\right)eV  \right)+g^K\left(\epsilon+\left(k-\frac{1}{2}\right)eV\right) -g^K\left(\epsilon-\left(k+\frac{1}{2}\right)eV  \right) \right.\nonumber\\ &&\left. -g^K\left(\epsilon-\left(k-\frac{1}{2}\right)eV \right)\right]e^{i\vartheta}\delta(2k-n)-
\left[f^K(\epsilon+keV) +  f^K(\epsilon-keV) \right]e^{i\vartheta}\nonumber\\&&\times\left[ \delta(2k+1-n)+\delta(2k-1-n) \right].
\end{eqnarray}
In addition, we employed the definitions
\begin{eqnarray}
&& \tilde F(x,q)=F_1\left(x-\frac{eV}{2}\right)+F_2\left(x+\frac{eV}{2}\right),\;{\rm if}\; q> 0,
\label{tilf}\\ && \tilde F(x,q)=F_1\left(x-\frac{eV}{2}\right)+F_1\left(x+\frac{eV}{2}\right),\;{\rm if}\; q=0,\nonumber
\\ && \tilde F(x,q)=F_1\left(x+\frac{eV}{2}\right)+F_2\left(x-\frac{eV}{2}\right),\;{\rm if}\; q<0\nonumber
\end{eqnarray}
and
\begin{eqnarray}
&& \widetilde{\widetilde{ F}}(x,m)=F_2\left(x+\frac{eV}{2}\right)-F_1\left(x-\frac{eV}{2}\right),\;{\rm if}\; m> 0,
\label{tiltilf}\\ && \widetilde{\widetilde F}(x,m)=F_1\left(x+\frac{eV}{2}\right)-F_1\left(x-\frac{eV}{2}\right),\;{\rm if}\; m=0,\nonumber
\\ && \widetilde{\widetilde{ F}}(x,m)=F_1\left(x+\frac{eV}{2}\right)-F_2\left(x-\frac{eV}{2}\right),\;{\rm if}\; m<0,\nonumber
\end{eqnarray}
where
\begin{eqnarray}
&& F_1(x)=\tanh\left( \frac{x}{2T}\right)\frac{1-\left| a^R\right|^2(x)}{1-\left(a^{R*}\right)^2(x)},\label{fexpr}
\\&& F_2(x)=-\tanh\left( \frac{x}{2T}\right)\frac{a^{R*}(x)}{a^R(x)}\frac{1-\left| a^R\right|^2(x)}{1-\left(a^{R*}\right)^2(x)}=-\frac{a^{R*}(x)}{a^R(x)}F_1(x).\nonumber
\end{eqnarray}
The functions $\tilde\zeta^{R,A}$ were already introduced in Appendix A. They read
\begin{equation}
\tilde\zeta^R\left(\epsilon+\frac{eV}{2},l \right)=\left\{ \begin{array}{l} (-1)^l\prod_{1\le k \le l}a^R(\epsilon+eVk),\quad {\rm if}\; l>0, \\ 1,\quad {\rm if}\; l=0, \\ (-1)^l\prod_{l+1\le k \le 0}a^R(\epsilon+eVk),\quad {\rm if}\; l<0, \end{array}\right.\label{iel}
\end{equation}
\begin{equation}
\tilde\zeta^A\left(\epsilon-\frac{eV}{2},l \right)=\left\{ \begin{array}{l} (-1)^l\prod_{1\le k \le l}a^{R*}(\epsilon-eVk),\quad {\rm if}\; l>0, \\ 1,\quad {\rm if}\; l=0 , \\ (-1)^l\prod_{l+1\le k \le 0}a^{R*}(\epsilon-eVk),\quad {\rm if}\; l<0. \end{array}\right.
\end{equation}

\section{Arbitrary microwave signals}

Provided the phase $\varphi(t)$ is defined in Eq. (\ref{acv}), we may write
\begin{eqnarray}
&& \cos\frac{ \varphi (t)- \varphi(t')}{4}=\frac{1}{2}e^{ieV(t-t')/2}\left[\sum_{k,l=-\infty}^{\infty}J_{k+l}\left(\frac{\alpha}{2} \right) J_{k-l}\left(\frac{\alpha}{2} \right) e^{ik\omega(t-t')+il\omega(t+t')}+\right.\label{be1}
\\ &&\left.+\sum_{k,l=-\infty}^{\infty}J_{k+l+1}\left(\frac{\alpha}{2} \right) J_{k-l}\left(\frac{\alpha}{2} \right) e^{i\left(k+\frac{1}{2}\right)\omega(t-t')+i\left(l+\frac{1}{2}\right)\omega(t+t')}\right]+c.c.\nonumber
\end{eqnarray}
Here the parameter $\alpha$ is not necessarily small. We also have
\begin{eqnarray}
\label{C2}
&& \cos\frac{ \varphi (t)+ \varphi (t')}{4}=\frac{1}{2} e^{i\vartheta+ieV(t+t')/2}\left[\sum_{k,l=-\infty}^{\infty}J_{k+l}\left(\frac{\alpha}{2} \right) J_{k-l}\left(\frac{\alpha}{2} \right) e^{ik\omega(t+t')+il\omega(t-t')}+\right. \label{be2}\\ && \left. +\sum_{k,l=-\infty}^{\infty}J_{k+l+1}\left(\frac{\alpha}{2} \right) J_{k-l}\left(\frac{\alpha}{2} \right) e^{i\left(k+\frac{1}{2}\right)\omega(t+t')+i\left(l+\frac{1}{2}\right)\omega(t-t')} \right]+c.c. \nonumber
\end{eqnarray}

Similarly to Eq. (\ref{feq}) let us define
\begin{equation}
x(t,t')=\sum_{n,m=-\infty}^\infty \int\frac{d\epsilon}{2\pi} x(\epsilon,n,m) e^{-i\epsilon(t-t')} e^{-i ne V(t+t')/2}e^{-im\omega(t+t')/2}.\label{feq2}
\end{equation}
For the convolution $z(t,t')=x(t,n,m)\circ y(t,k,l)$ we then have
\begin{equation}
z(\epsilon,n+k,m+l)=x\left(\epsilon+\frac{keV}{2}+\frac{l\omega}{2},n,m\right)y\left(\epsilon-\frac{neV}{2}
-\frac{m\omega}{2},k,l\right).\label{crule3}
\end{equation}
The relationship defining the inverse matrix $\hat X^R$ can be expressed in the form
\begin{equation}
\left( \hat a^R\circ \hat X^R\right)\left( \epsilon,n,m\right)=\delta_{n,0}\delta_{m,0},\label{invmax}
\end{equation}
where for even and odd values of $m$ we have respectively
\begin{eqnarray}
&& \hat a^R(\epsilon,n,m)=\label{ashae}\\&&\frac{1}{2}\delta_{n,0}\sum_{k}J_{k+\frac{m}{2}}\left( \frac{\alpha}{2}\right)J_{k-\frac{m}{2}}\left( \frac{\alpha}{2}\right)\left[g^R \left(\epsilon+\frac{eV}{2}-k\omega \right)+g^R \left(\epsilon-\frac{eV}{2}-k\omega \right)\right] \left(\begin{array}{cc} 1& 0\\ 0 & 1\end{array}\right)+\nonumber
\\ && \frac{1}{2}\left( \delta_{n,1}e^{-i\vartheta}+\delta_{n,-1}e^{i\vartheta}\right)\sum_l J_{\frac{m}{2}+l}\left( \frac{\alpha}{2}\right)J_{\frac{m}{2}-l}\left( \frac{\alpha}{2}\right)f^R(\epsilon-l\omega)\left(\begin{array}{cc} 0& 1\\ 1 & 0\end{array}\right)\nonumber
\end{eqnarray}
and
\begin{eqnarray}
&& \hat a^R(\epsilon,n,m)=\frac{1}{2}\delta_{n,0}\sum_{k}J_{k+\frac{1}{2}+\frac{m}{2}}\left( \frac{\alpha}{2}\right)J_{k+\frac{1}{2}-\frac{m}{2}}\left( \frac{\alpha}{2}\right)\times\label{ashao}\\&&\times\left[-g^R \left(\epsilon+\frac{eV}{2}-\left(k+\frac{1}{2} \right)\omega \right)+g^R \left(\epsilon-\frac{eV}{2}-\left(k+\frac{1}{2} \right)\omega \right)\right] \left(\begin{array}{cc} 1& 0\\ 0 & 1\end{array}\right)+\nonumber
\\ && \frac{1}{2}\left( \delta_{n,1}e^{-i\vartheta}-\delta_{n,-1}e^{i\vartheta}\right)\sum_l J_{\frac{m}{2}+l+\frac{1}{2}}\left( \frac{\alpha}{2}\right)J_{\frac{m}{2}-l-\frac{1}{2}}\left( \frac{\alpha}{2}\right)f^R\left(\epsilon-\left( l+\frac{1}{2}\right)\omega\right)\left(\begin{array}{cc} 0& 1\\ 1 & 0\end{array}\right).\nonumber
\end{eqnarray}

As before, we define
\begin{equation}
\hat X^R(\epsilon,n,m)=\left(\begin{array}{cc} \beta^R(\epsilon,n,m) & \gamma^R(\epsilon,n,m)\\ \gamma^R(\epsilon,n,m)& \beta^R(\epsilon,n,m) \end{array} \right),
\end{equation}
where $\beta^R$ differs from zero for even $n$ whereas $\gamma^R$ is non-zero for odd $n$. We again introduce
\begin{equation}
\zeta^R(\epsilon,n,m)=\left\{ \begin{array}{c} \beta^R(\epsilon,n,m)\quad {\rm for\; even}\; n,\\ \gamma^R(\epsilon,n,m)\quad {\rm for\; odd}\; n,\end{array}\right. \quad   \tilde\zeta^R (\epsilon,n,l)= \zeta^R \left(\epsilon+\frac{neV}{2}+\frac{l\omega}{2},n,l\right)
\label{zint}
\end{equation}
and compose the column
\begin{equation}
\hat \zeta^R(\epsilon)=\left(\begin{array}{c} \vdots\\ \tilde \zeta^R(\epsilon,n-1,...)\\ \tilde \zeta^R(\epsilon,n,...)\\ \tilde \zeta^R(\epsilon,n+1,...)\\ \vdots\end{array} \right),
\end{equation}
where dots in $\tilde \zeta^R(\epsilon,n,...)$ indicate that the index $l$ is running in its range.

The system of equations (\ref{invmax}) can then be expressed in the form
\begin{equation}
\hat T\hat\zeta^{R}=\delta_{k,0}\delta_{n,0},\label{geq}
\end{equation}
where the matrix $\hat T$ has the block tridiagonal structure
\begin{equation}
\hat T=\left( \begin{array}{ccccc}  \hat D_1 & -e^{i\vartheta} \hat A_2^T &\phantom{d} &\phantom{d} & \phantom{d}\\ -e^{-i\vartheta} \hat A_2 &  \hat D_2 & -e^{i\vartheta} \hat A_3^T &\phantom{d} & \phantom{d} \\ \phantom{d} & \ddots &\ddots &\ddots & \phantom{d}
\\ \phantom{d} & \phantom{d} & -e^{-i\vartheta} \hat A_{N-1}&  \hat D_{N-1} & -e^{i\vartheta}  \hat A_N^T  \\ \phantom{d} & \phantom{d} & \phantom{d}& -e^{-i\vartheta} \hat A_N &  \hat D_N\end{array}\right)
\nonumber
\end{equation}
similar to that discussed in Ref. \onlinecite{M} and the indices $k$ and $n$ again correspond respectively to frequency and voltage harmonics.
The factors $e^{\pm i\vartheta}$ factors can be eliminated by means of a unitary transformation
\begin{equation}
\hat T=\hat U^\dagger \hat T_0 \hat U, \quad  \hat T_0=\left( \begin{array}{ccccc} \hat D_1 & -\hat A_2 &\phantom{d} &\phantom{d} & \phantom{d}\\ - \hat A_2 &\hat D_2 &- \hat A_3 &\phantom{d} & \phantom{d} \\ \phantom{d} & \ddots &\ddots &\ddots & \phantom{d}
\\ \phantom{d} & \phantom{d} & - \hat A_{N-1}& \hat D_{N-1} &  -\hat A_N  \\ \phantom{d} & \phantom{d} & \phantom{d}& - \hat A_N & \hat D_N\end{array}\right)\label{unt}
\end{equation}
with $\hat U$ being a diagonal unitary matrix with elements $U_{nn}=e^{i\phi_n}$. Equation (\ref{unt}) holds for $\phi_{n+1}-\phi_{n}=\vartheta$.
Provided the difference $k-l$ is even, for the elements $D^{\left[ n\right]}_{kl}(\epsilon)$ of the matrix $ \hat D_n$ we get
\begin{eqnarray}
&&D^{\left[ n\right]}_{kl}(\epsilon)=\frac{1}{2}\sum_{s=-\infty}^\infty J_{s+\frac{k-l}{2}}\left(\frac{\alpha}{2} \right)J_{s+\frac{l-k}{2}}\left(\frac{\alpha}{2} \right)\left[ g^R\left(\epsilon+\left(n+\frac{1}{2}\right)eV+\left(\frac{l+k}{2}-s\right)\omega \right)+\right.\nonumber\\ &&+\left.g^R\left(\epsilon+\left(n-\frac{1}{2}\right)eV+\left(\frac{l+k}{2}-s\right)\omega \right)\right],\label{d1e}
\end{eqnarray}
whereas for odd values of $k-l$ we obtain
\begin{eqnarray}
&&D^{\left[ n\right]}_{kl}(\epsilon)=\frac{1}{2}\sum_{s=-\infty}^\infty J_{s+\frac{1+k-l}{2}}\left(\frac{\alpha}{2} \right)J_{s+\frac{1+l-k}{2}}\left(\frac{\alpha}{2} \right)\left[ g^R\left(\epsilon+\left(n-\frac{1}{2}\right)eV+\left(\frac{l+k-1}{2}-s\right)\omega \right)\right.\nonumber\\ &&\left.-g^R\left(\epsilon+\left(n+\frac{1}{2}\right)eV+\left(\frac{l+k-1}{2}-s\right)\omega \right)\right].\label{d1o}
\end{eqnarray}
The above equations indicate that the symmetry relation $D^{\left[ n\right]}_{kl}(\epsilon)=D^{\left[ n\right]}_{lk}(\epsilon)$ is obeyed.

Likewise, the elements $A^{\left[ n\right]}_{kl}(\epsilon)$ of the matrix $ \hat A_n$ respectively for even and odd values of the difference $k-l$ read
\begin{equation}
A^{\left[ n\right]}_{kl}(\epsilon)=-\frac{1}{2}\sum_{s} J_{\frac{k-l}{2}+s}\left(\frac{\alpha}{2}\right)J_{\frac{k-l}{2}-s} \left(\frac{\alpha}{2}\right) f^R\left(\epsilon+\left(n-\frac{1}{2}\right)eV+\left(\frac{l+k}{2}-s\right)\omega \right)\label{d2e}
\end{equation}
and
\begin{equation}
A^{\left[ n\right]}_{kl}(\epsilon)=-\frac{1}{2}\sum_{s} J_{\frac{k-l+1}{2}+s}\left(\frac{\alpha}{2}\right)J_{\frac{k-l-1}{2}-s} \left(\frac{\alpha}{2}\right) f^R\left(\epsilon+\left(n-\frac{1}{2}\right)eV+\left(\frac{l+k-1}{2}-s\right)\omega \right).\label{d2o}
\end{equation}
These equations imply that the matrix $A^{\left[ n\right]}_{kl}$ is symmetric for even $k-l$  and antisymmetric for odd $k-l$.
\end{widetext}

\section{Estimate for fractional Shapiro steps}

The matrix $\check Q_0^{-1}\equiv \check X$ defined in Eq. (\ref{cX}) can be identically rewritten as
\begin{equation}
\check X=\check X_0+\sum_{m=1}^\infty (-1)^m \check X_0\left(\delta \check a \check X_0 \right)^m.
\label{mappinv}
\end{equation}
Here $\check X_0$ is the inverse matrix  $\check Q_0^{-1}$ evaluated for $\check a_0=\delta (t-t')$ and $\delta \check a=\check a-\check a_0$.

In order to illustrate the idea of our estimate let us consider one of the terms in Eq. (\ref{abstcu}), e.g., the term containing the combination $\tau_3\hat X^R \tau_3\hat b^K$. One of the contributions to the current  (proportional to $\delta \check a$) generated by this term takes the form
(cf. the first term in the square brackets in Eq. (\ref{C2}))
\begin{eqnarray}
\nonumber
\frac{ie{\mathcal N}}{4}\int d t' f^K(t-t') f^R(t'-t)e^{ieV(t+t')+i\vartheta}\\
\times \sum_{p,l=-\infty}^{\infty}J_{p+l}\left(\frac{\alpha}{2} \right) J_{p-l}\left(\frac{\alpha}{2} \right) e^{ip \omega(t+t')+il\omega(t'-t)}
\label{D2}
\end{eqnarray}
Under the condition $eV=-p\omega$ the dependence of the combination \eqref{D2} on $t+t'$ drops out and it depends only on the time difference $t-t'$. Performing the Fourier transformation we receive the contributions containing the integrals
\begin{equation}
\sim \sum_l\int \frac{d\epsilon}{2\pi} f^K(\epsilon) f^R(\epsilon+l\omega)\label{estint}
\end{equation}
It is easy to see that the main contribution to this sum is provided by the term with $l=0$. In order to demonstrate that let us consider the combination
\begin{eqnarray}
\nonumber
&&\frac{ie{\mathcal N}}{4}\int\limits_{-\infty}^\infty \frac{d\epsilon}{2\pi} f^K(\epsilon) \left( f^R(\epsilon)+f^A(\epsilon)\right)\\
&=&\frac{ie{\mathcal N\Delta^2}}{4} \int\limits_{-\infty}^\infty  \frac{d\epsilon}{2\pi}\tanh\frac{\epsilon}{2T}\left(\mathcal{R}_{+}^{-2}(\epsilon) - \mathcal{R}_{-}^{-2}(\epsilon) \right)
\label{appeq}
\end{eqnarray}
entering into the expression for the current. Keeping track of a small imaginary part $\pm i\gamma$ in the expressions for ${\mathcal{R}_{\pm}(\epsilon)=\sqrt{(\epsilon\pm i\gamma)^2-\Delta^2}}$ and making use of the Sokhotski-Plemelj theorem,
for the combination (\ref{appeq}) we obtain
\begin{eqnarray}
\nonumber
&&e{\mathcal N}\Delta^2 \int\limits_{-\infty}^\infty  \frac{d\epsilon}{2\pi} \tanh\frac{\epsilon}{2T} \frac{\gamma \epsilon}{\left[ (\epsilon+i\gamma)^2-\Delta^2\right]\left[ (\epsilon-i\gamma)^2-\Delta^2\right] }\\
&=&\frac{e{\mathcal N}\Delta^2}{8} \int\limits_{-\infty}^\infty  \frac{d\epsilon}{\epsilon} \tanh\frac{\epsilon}{2T}
\left[\delta(\epsilon+\Delta)+ \delta(\epsilon-\Delta) \right]
\nonumber\\
&=&\frac{e{\mathcal N}\Delta}{4} \tanh\frac{\Delta}{2T}\sim I_c.
\label{D5}
\end{eqnarray}
Thus we observe that the main contribution to the integral with $l=0$ in Eq. (\ref{estint}) comes from $\epsilon=\pm \Delta$ since
the expression under this integral strongly peaked at these two values of $\epsilon$. The terms with $l \neq 0$ contain no such peaks, hence, their contributions are smaller and can be safely neglected.

Substituting the estimate \eqref{D5} into Eq. \eqref{D2} and identifying $n=-2p$ we arrive at the result for the magnitude of fractional Shapiro steps $\delta I_{\frac {1}{n}}$ which occur at $\omega=\omega_J/n$ with even $n$:
\begin{equation}
\delta I_{\frac {1}{n}}\sim I_c J^2_{n/2}(\alpha /2).
\label{D6}
\end{equation}

Repeating the whole analysis with the second term in the square brackets of Eq. \eqref{C2} we recover the analogous estimate for odd values of $n$:
\begin{equation}
\delta I_{\frac {1}{n}}\sim I_c J_{(n+1)/2}(\alpha)  J_{(n-1)/2}(\alpha /2).
\label{D7}
\end{equation}

Finally, we note that the above simple analysis includes only linear in $\delta \check a$ terms in the formal series \eqref{mappinv}.  Hence, one may expect that the estimates \eqref{D6} and \eqref{D7} should work reasonably well for $\alpha \lesssim 1$ and may become less accurate for bigger values of $\alpha$.


\begin{thebibliography}{99}
\bibitem{FN2} This conclusion is fully consistent, e.g., with the observation of a fairly large excess current in the experiments \cite{M16,topmat,Gre}.
\end{thebibliography}
\end{document}